\documentclass[preprint,numberedappendix,appendixfloats,twocolappendix]{emulateapj}

\usepackage{subfigure}
\usepackage{graphicx}
\usepackage[usenames]{color}
\usepackage{enumitem}

\received{}
\revised{}
\accepted{}

\shortauthors{Boyer et al.}
\shorttitle{``DUSTiNGS Paper II''}

\begin{document}

\title{An Infrared Census of DUST In Nearby Galaxies with Spitzer
  (DUSTiNGS), II. Discovery of Metal-poor Dusty AGB Stars}

\author{Martha~L.~Boyer\altaffilmark{1,2},
  Kristen~B.~W. McQuinn\altaffilmark{3}, Pauline
  Barmby\altaffilmark{4}, Alceste~Z. Bonanos\altaffilmark{5}, Robert
  D. Gehrz\altaffilmark{3}, Karl~D.~Gordon\altaffilmark{6},
  M.~A.~T. Groenewegen\altaffilmark{7}, Eric Lagadec\altaffilmark{8},
  Daniel Lennon\altaffilmark{9}, Massimo Marengo\altaffilmark{10},
  Iain McDonald\altaffilmark{13}, Margaret~Meixner\altaffilmark{6},
  Evan Skillman\altaffilmark{3}, G.~C.~Sloan\altaffilmark{11},
  George~Sonneborn\altaffilmark{1},
  Jacco~Th.~van~Loon\altaffilmark{12},
  Albert~Zijlstra\altaffilmark{13}} 
\altaffiltext{1}{Observational
  Cosmology Lab, Code 665, NASA Goddard Space Flight Center,
  Greenbelt, MD 20771 USA; martha.boyer@nasa.gov} 
\altaffiltext{2}{Oak
  Ridge Associated Universities (ORAU), Oak Ridge, TN 37831 USA}
\altaffiltext{3}{Minnesota Institute for Astrophysics, School of Physics and Astronomy, 116 Church Street SE, University of
  Minnesota, Minneapolis, MN 55455 USA}  
\altaffiltext{4}{Department of Physics \& Astronomy, University of Western Ontario, London, ON, N6A 3K7, Canada}
\altaffiltext{5}{IAASARS, National Observatory of Athens, GR-15236 Penteli, Greece}
\altaffiltext{6}{STScI, 3700 San Martin Drive, Baltimore, MD 21218
  USA}
\altaffiltext{7}{Royal Observatory of Belgium, Ringlaan 3, B-1180 Brussels, Belgium}
\altaffiltext{8}{Laboratoire Lagrange, UMR7293, Univ. Nice Sophia-Antipolis, CNRS, Observatoire de la C\^{o}te d'Azur, 06300 Nice, France}
\altaffiltext{9}{ESA - European Space Astronomy Centre, Apdo. de Correo 78, 28691 Villanueva de la Ca\~{n}ada, Madrid, Spain}
\altaffiltext{10}{Department of Physics and Astronomy, Iowa State University, Ames, IA 50011, USA}
\altaffiltext{11}{Astronomy Department, Cornell University, Ithaca, NY 14853-6801, USA}
\altaffiltext{12}{Astrophysics Group, Lennard-Jones Laboratories,
  Keele University, Staffordshire ST5 5BG, UK}
\altaffiltext{13}{Jodrell Bank Centre for Astrophysics, Alan Turing Building, University of Manchester, M13 9PL, UK}

\begin{abstract}
The DUSTiNGS survey (DUST in Nearby Galaxies with {\it Spitzer}) is a
3.6 and 4.5 \micron\ imaging survey of 50 nearby dwarf galaxies
designed to identify dust-producing Asymptotic Giant Branch (AGB)
stars and massive stars.  Using 2 epochs, spaced approximately 6
months apart, we identify a total of 526 dusty variable AGB stars
(sometimes called ``extreme'' or x-AGB stars;
$[3.6]-[4.5]>0.1$~mag). Of these, 111 are in galaxies with ${\rm
  [Fe/H]} < -1.5$ and 12 are in galaxies with ${\rm [Fe/H]} < -2.0$,
making them the most metal-poor dust-producing AGB stars known. We
compare these identifications to those in the literature and find that
most are newly discovered large-amplitude variables,
with the exception of $\approx$30 stars in NGC\,185 and NGC\,147, 1
star in IC\,1613, and 1 star in Phoenix.  The chemical abundances of
the x-AGB variables are unknown, but the low metallicities suggest
that they are more likely to be carbon-rich than oxygen-rich and
comparisons with existing optical and near-IR photometry confirms that
70 of the x-AGB variables are confirmed or likely carbon stars. We see
an increase in the pulsation amplitude with increased dust production,
supporting previous studies suggesting that dust production and
pulsation are linked. We find no strong evidence linking dust
production with metallicity, indicating that dust can form in very
metal-poor environments.

\end{abstract}


\vfill\eject
\section{INTRODUCTION}
\label{sec:intro}

The origin of the massive dust reservoirs in high-redshift, metal-poor
quasars is under dispute \citep[$z\lesssim6$, $M>10^{8}
  M_{\odot}$;][]{Bertoldi+2003, Robson+2004, Beelen+2006}. Some
authors argue that core-collapse supernovae (SNe) are the dominant
stellar dust source \citep[e.g.,][]{Michalowski+2010}, while others
argue that Asymptotic Giant Branch (AGB) stars contribute
significantly \citep[e.g.,][]{Valiante+09,Zhukovska+2013}.
Progenitors of core-collapse SNe produce the seeds necessary for dust
production in their cores and are thus capable of producing dust at
any metallicity, though it is unclear how much of the dust they
produce will survive the reverse shock
\citep[e.g.,][]{Dwek+08,Kozasa+2009}. In AGB stars, the
dust-production process is not well understood.  AGB stars can be
carbon- or oxygen-rich, depending on how much free oxygen is left
after the dredge up of newly formed carbon; metal-poor AGB stars are
more likely to be C-rich since less oxygen is available. These stars
produce carbon in situ that can condense into amorphous carbon grains,
but it is unclear whether the dust condensation process is limited by
the metallicity of the star.  AGB mass-loss models from
\citet{Wachter+2008} find that dust-driven mass-loss rates for solar
and sub-solar metallicity carbon (C) stars are similar (down to
$Z=0.001$), but that the metal-poor stars make less dust
overall. Mattsson et al. (in preparation) argue that metal-poor C
stars are not a major dust source, except perhaps at the very end of
their evolution.

While very dusty C stars have been observed in the Magellanic Clouds
(MCs) at ${\rm [Fe/H]} \gtrsim -1$
\citep[e.g.,][]{GroenewegenBlommaert1998,vanLoon+97,vanLoon+99,vanLoon+08b,Zijlstra+06,Groenewegen+07,Lagadec+07,Gruendl+2008,Riebel+2012},
there are few observations of similar stars at lower
metallicities. Searches in nearby dwarf galaxies yield only 2 dusty C
stars in Sculptor and 3 in Leo\,I \citep[${\rm [Fe/H]} = -1.68$ and
  $-1.43$, respectively;][]{Sloan+2012}, though these examples produce
less dust (by more than an order of magnitude) than what is seen in MC
stars. Using statistical arguments to estimate the size of the AGB
population in several dwarf irregular galaxies (down to ${\rm [Fe/H]}
= -2.1$), \citet{Jackson+07a,Jackson+07b} and \citet{Boyer+09b} find
evidence for the presence of dust-producing stars, but were unable to
identify individual stars owing to confusion with unresolved
background galaxies. In this work, we identify hundreds of dusty AGB
star candidates in galaxies more metal-poor than the MCs and
investigate the effect of metallicity on dust production. The results
have important implications for the dust budgets of high-redshift,
metal-poor galaxies.

\subsection{Extreme (x-)AGB Stars}
\label{xAGB_intro}

The Surveying the Agents of Galaxy Evolution (SAGE) {\it Spitzer}
surveys of the Small and Large Magellanic Clouds (SMC/LMC) obtained
3.6--70~\micron\ photometry covering the full spatial extent of each
galaxy, thereby detecting the circumstellar dust around the
near-complete population of AGB stars in each galaxy
\citep{Meixner+06,Gordon+11}.  In both galaxies, a subset of AGB stars
with colors $J-K \gtrsim 2$~mag show evidence for significantly more
dust production than that observed for the average AGB star. Most of
these stars are C-rich \citep{Woods+2011}, and their mass-loss rates
exceed the nuclear-burning mass consumption rate
\citep[cf.][]{Boyer+2012}. These stars are sometimes referred to as
``extreme'' (x-)AGB stars, and we use this terminology here to be
consistent with the recent work in the MCs \citep[e.g.,][]{Blum+06}.

The x-AGB stars comprise $<$6\% of the total AGB population, but they
account for more than 75\% of the dust produced by cool evolved stars
\citep{Matsuura+09,Srinivasan+2009,Boyer+2012,Riebel+2012}. These
stars also contribute significantly to the global 8-\micron\ flux of
the SMC \citep{MelbourneBoyer2013}.  Other than those in the MCs and
the Milky Way, x-AGB stars have been detected and their dust
production confirmed in only a few galaxies: M33
\citep{McQuinn+2007,Javadi+2013}, M32 \citep{Jones+2014}, Sgr dSph
\citep{McDonald+2012}, Fornax and Sculptor
\citep{Matsuura+2007,Groenewegen+09b,Sloan+09}, so it is unclear
whether AGB stars in metal-poor galaxies are efficient dust producers.

\subsection{DUSTiNGS}
\label{sec:dustings}

\citet[][hereafter Paper\,I]{Boyer+2014} describe the Dust in Nearby
Galaxies with {\it Spitzer} (DUSTiNGS) survey in detail. DUSTiNGS is a
3.6 and 4.5 \micron\ imaging survey of 50 nearby dwarf galaxies
designed to detect all dust-producing evolved stars.  Targets include
37 dwarf spheroidal (dSph), 8 dwarf irregular (dIrr), and 5 transition
type (dTrans or dSph/dIrr) galaxies (see Table~1 from Paper\,I). The
target galaxies have experienced a variety of different star formation
and interaction histories and span a wide range in mass traced by
visible light ($0 > M_{\rm V} > 15.2$~mag) and metallicity ($-2.72 <
{\rm [Fe/H]} < -1.1$).

It is difficult to separate AGB stars belonging to a galaxy from
unresolved background sources and foreground stars at these
wavelengths. In Paper\,I, we used the large field of view to estimate
the size of the thermally-pulsing (TP-)AGB population by statistically
estimating and removing the foreground and background sources for each
galaxy, assuming that the tip of the Red Giant Branch (TRGB) is at
$M_{3.6} = -6$~mag. We found that the Andromeda satellites in the
sample harbor $40 \pm 30$ to $506 \pm 48$ AGB stars each (including
dusty and non-dusty stars), with the largest populations in And\,VII,
And\,XVIII, and And\,X. The most metal-poor galaxies in the sample
have the lowest masses, and therefore are the least likely to harbor
AGB stars. In these galaxies, we found only upper limits on the size
of the AGB population ($<$9 each), with the exception of $20 \pm 9$
AGB stars in the Hercules Dwarf (${\rm [Fe/H]} = -2.41$). The largest
AGB populations are in the star-forming dIrr galaxies.

In Paper\,I, we also found evidence for x-AGB stars in 8 of the
DUSTiNGS galaxies (And\,II, Cetus, IC\,10, IC\,1613, NGC\,147,
NGC\,185, Sextans\,B, and WLM), though it is impossible to isolate
galaxy members from red background sources with only 3.6 and
4.5~\micron\ imaging. In order to identify individual x-AGB stars, the
DUSTiNGS images were obtained in 2 epochs, separated by approximately
6 months.  Dust-producing AGB stars are expected to pulsate with
periods ranging from 100--1000~days \citep{VassiliadisWood93}, though
little is known about how the mid-IR properties of these stars are
affected by pulsation.  Light curves for pulsating AGB stars in the
MCs, the Milky Way, and in nearby dwarf spheroidal galaxies show that
pulsation amplitudes decrease from optical to near-infrared (IR)
wavelengths
\citep{LeBertre1992,LeBertre1993,Whitelock+2003,Menzies+2010,BattinelliDemers2012},
but 1-20~\micron\ light curves of Galactic AGB stars from
\citet{Harvey+1974} and \citet{LeBertre1992,LeBertre1993} suggest that
amplitudes of dusty sources may increase in the IR owing to changes in
the warm circumstellar dust that is responding to the pulsations of
the photosphere. \citet{McQuinn+2007} obtained 5 epochs of imaging of
M33 with {\it Spitzer}, and were able to identify a large population
of dust-producing C star candidates with 3.6~\micron\ amplitudes up to
0.8~mag (amplitudes in that work are the standard deviation of over the
5 epochs).  The SAGE observations of the MCs obtained 2 epochs of
imaging, and find that the x-AGB stars show amplitudes up to 1.4~mag
at 3.6~\micron\ \citep{Vijh+09,Polsdofer+2014}.

Here, we report the identification of 526 x-AGB star candidates in the
DUSTiNGS galaxies that were discovered via their 3.6 and
4.5~\micron\ variability, including 12 in galaxies with ${\rm [Fe/H]}
< -2$. These are the most metal-poor dust-producing AGB stars known
thus far. This paper is organized as follows. In
Section~\ref{sec:analysis}, we describe the data, stellar
classification, and variability analysis. In Section~\ref{sec:known},
we compare the DUSTiNGS variables to previously detected variables and
C stars in these galaxies. In Section~\ref{sec:disc}, we discuss the
spatial distributions, amplitudes, and dust production of the x-AGB
star candidates. We summarize the results in Section~\ref{sec:concl}.

\section{Data \& Observations}
\label{sec:analysis}

To identify candidate variable AGB stars, we use the 2-epoch DUSTiNGS
data and an additional epoch of data from the {\it Spitzer} archive
from earlier programs. We use the {\it Spitzer} colors to identify
variable stars that are producing dust.

\subsection{DUSTiNGS Data}
\label{sec:data}

The DUSTiNGS observations and photometry are described in detail in
Paper\,I. In brief, each galaxy was imaged simultaneously at 3.6 and
4.5~\micron\ to at least the half-light radius ($r_{\rm h}$) in 2
epochs. The separation between epochs was determined by {\it
  Spitzer's} visibility windows for each target.  The average epoch
separation is 180 days, with a range of 127--240 days (Table~2 from
Paper\,I). All photometry that we use here is from the ``good-source''
catalog (GSC), which has been culled to include only high-confidence
point sources and reliable measurements (Paper\,I).

Padova stellar evolution models suggest that $>$90\% of the TP-AGB
population is brighter than the TRGB at a given time \citep[][and
  G. Bruzual, private
  communication]{Marigo+08,Marigo+13}. The TRGB ranges from
$M_{3.6} = -6.6$ to $-6$~mag, based on {\it Spitzer} observations of 8
dIrr galaxies \citep{Jackson+07a,Jackson+07b,Boyer+09b}. To ensure
that the majority of the TP-AGB stars are measured, the DUSTiNGS
photometry is at least 75\% complete down to this limit. However, the
IRAC colors and magnitudes of AGB stars make it difficult to
distinguish them from foreground stars and unresolved background
galaxies. This is especially true for target galaxies more distant
than $\approx$250~kpc ($(m-M)_0 \gtrsim 22$~mag), for which individual
AGB stars can be fainter than unresolved background galaxies. In Paper
I, we estimated the total AGB population size by statistically
subtracting the background and foreground contamination. Here, we use
stellar variability to identify a subset of individual AGB candidates.

\subsection{Additional Spitzer Data}
\label{sec:cryo}

Eight of the DUSTiNGS targets were also observed with {\it Spitzer}
during the cryogenic mission. The observations for these galaxies are
described in detail in \citet{Jackson+07a,Jackson+07b} and
\citet{Boyer+09b}. In brief, the total integration times and dithering
strategies are nearly identical to DUSTiNGS, but the fields of view
are significantly smaller.  For 6 galaxies (Aquarius, Leo\,A, LGS\,3,
Pegasus, Phoenix, and Sextans\,A), only a single
$\approx$5\arcmin$\times$5\arcmin\ IRAC frame was observed.  IC\,1613
was imaged in a 2$\times$3 IRAC map and WLM in a 3$\times$1 IRAC map
(See Figs.~\ref{fig:spat2} and \ref{fig:spat3}).  We downloaded the
data from these programs (Programs 128 and 40524, P.Is.: R. D. Gehrz
and E. D. Skillman, respectively), processed with the S18.25.0
pipeline, from the {\it Spitzer} Heritage Archive and produced
photometric catalogs following the same process used for the DUSTiNGS
data (Paper\,I).

The 3.6~\micron\ observations for all 8 galaxies and both 3.6 and
4.5~\micron\ observations for IC\,1613 and WLM were obtained prior to
2006 (Program 128). The 4.5~\micron\ observations for the remaining 6
galaxies were obtained in 2007 (Program 40524).  See
\citet{Jackson+07a,Jackson+07b} and \citet{Boyer+09b} for specific
dates for each set of observations. Since this epoch occurs prior to
the DUSTiNGS observations, we refer to it as `epoch 0'.

\section{Identifying AGB Stars}

We use variability between the epochs to identify
  individual AGB stars and use colors and magnitudes to separate x-AGB
  stars from the less dusty AGB stars. Tables~\ref{tab:vars} and
  \ref{tab:vars3} show the total numbers of variable AGB stars in each
  galaxy and Table~\ref{tab:varcandy} includes the resulting list of
  classified variables.

\subsection{Color Classification}
\label{sec:class}

%%%%%%%%%%%%%%%%%%%%%%%%%%%%%%%%%%%%%%%%%%%%%%%%%%%%%%%%%%%%%%%
\begin{figure}
\includegraphics[width=\columnwidth]{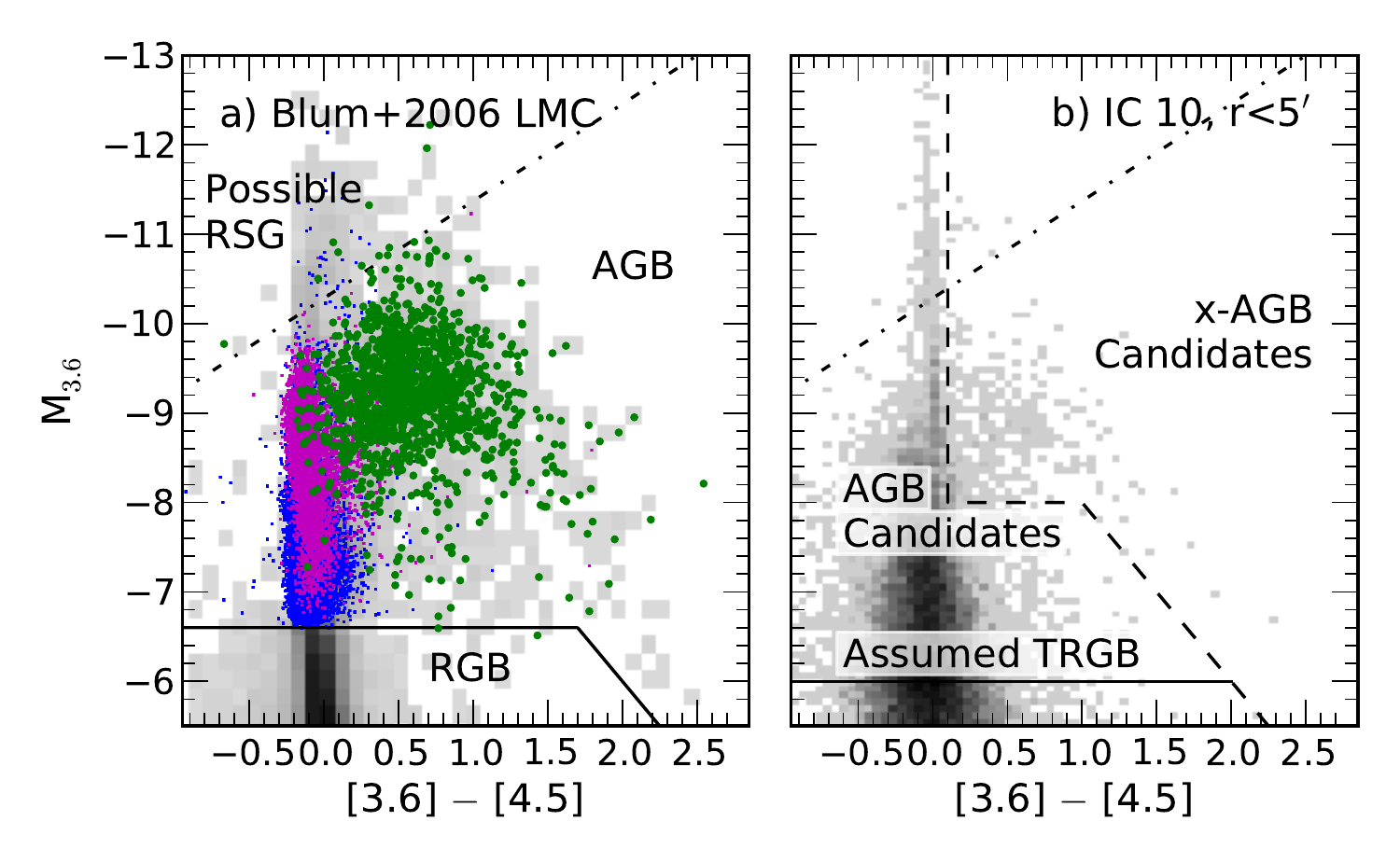}
\caption{Classification of Variable Sources. {\it (a)} LMC Evolved
  Stars from \citet{Blum+06}. Blue and magenta dots are
    less dusty oxygen-and carbon-rich AGB stars, respectively. Green
    dots are x-AGB stars. The greyscale includes all evolved
  stars, including red supergiants (RSG), AGB stars, and red giant
  branch (RGB) stars.  Young main sequence stars are not included, but
  are expected to have $[3.6]-[4.5]=0$~mag. The dashed lines show the
  approximate boundaries between RSG, AGB, and RGB stars. In the LMC,
  the TRGB is at $-6.6$~mag \citep{Boyer+11}. {\it (b)} IC\,10 CMD
  within $r=5\arcmin$ ($r_{\rm h} = 2\farcm65$). The TRGB is unknown
  for most DUSTiNGS galaxies, so we assume it lies at $-6$~mag. To
  mimic the classification scheme used in the LMC, we classify variable
  sources found to the right of the dashed line as x-AGB candidates.
  Variable sources to the left are AGB candidates, and those below the
  assumed TRGB are classified as ``unknown''. \label{fig:class}}
\end{figure}
%%%%%%%%%%%%%%%%%%%%%%%%%%%%%%%%%%%%%%%%%%%%%%%%%%%%%%%%%%%%%%%

In the LMC, \citet{Blum+06} designate stars with colors $J-[3.6]>
3.1$~mag as x-AGB stars. This color is empirically observed as the
point where there is a sudden break, or decrease in source density, in
the C star sequence evident in a $J-[3.6]$ vs. $[3.6]$ color-magnitude
diagram. This break roughly corresponds to the superwind phase of AGB
evolution, wherein the mass-loss rate increases dramatically due to an
increase in dust production by a factor of 10 or more
\citep{Schroder+1999}. At low metallicities, the superwind phase may
be triggered by the dredge-up of carbon
\citep{LagadecZijlstra2008}. Many AGB stars not classified as x-AGB
also show dust emission, though the dust emission from x-AGB stars is
significantly stronger.

Most of the DUSTiNGS galaxies do not have available $J$-band
photometry at the necessary depth to use the \citet{Blum+06}
classification scheme.  Instead, we mimic it based on the positions of
AGB stars in the $[3.6]-[4.5]$ CMD (Fig.~\ref{fig:class}a) and classify
variables as follows:

\begin{description}[leftmargin=1em]

\item[Unknown] Variables fainter than the assumed TRGB
  ($M_{3.6}=-6$~mag; solid line in Fig.~\ref{fig:class}b). The TRGB is
  not known for most of the DUSTiNGS galaxies, but it can be as bright
  as $M_{3.6} = -6.6$ \citep{Jackson+07a,Jackson+07b,Boyer+09b}.
\item[Less dusty AGB stars] Variables brighter than the assumed
  TRGB and bluer than the dashed line in Figure~\ref{fig:class}b.
\item[x-AGB stars] Variables redder than the dashed line in
  Figure~\ref{fig:class}b. The luminosity limit for x-AGB stars
  becomes fainter at redder colors to allow for the obscuration of the
  dustiest sources while also avoiding the region dominated by
  unresolved background sources (see Paper\,I).
\item[Red supergiants] AGB stars can exceed the classical AGB limit
  when experiencing hot-bottom burning \citep{Boothroyd+92}, so
  variables brighter than the dot-dashed line in
  Figure~\ref{fig:class} could be either massive red supergiants
  (RSGs) or AGB stars.  We classify these stars as AGB or x-AGB
  depending on their color and also flag them as potential RSGs in
  Table~\ref{tab:varcandy}.

\end{description}

Our classification scheme would correctly classify 93--94\% of the
\citet{Blum+06} x-AGB stars in the LMC and SMC. Similarly, $<$2\% of
other AGB types in the MCs would be mis-classified as x-AGB using our
scheme.  

We note that the x-AGB stars here are not the same as the very extreme
AGB stars in the LMC, as defined by \citet{Gruendl+2008}. That work
discovered 13 LMC C stars that are among the reddest objects in
that galaxy ($[3.6]-[4.5]$ up to 3.4~mag). All of the
\cite{Gruendl+2008} stars are so dust enshrouded that they are faint
even at 3.6~\micron\ ($M_{3.6} > -5$~mag), and are beyond the
sensitivity limit of DUSTiNGS. Only one star in the DUSTiNGS sample
approaches the colors of the \citet{Gruendl+2008} sources: a star in
IC\,10 with $[3.6]-[4.5] = 2.4$~mag (see Section~\ref{sec:cmd} and
Fig.~\ref{fig:class}).

Unknown variable stars may be AGB stars in the minimum brightness
phase of pulsation, but they could also be non-AGB variable sources
such as RR\,Lyrae, eclipsing binaries, or background Active Galactic
Nuclei (AGN). Cepheid variables are too faint at these wavelengths, so
we do not expect them to contaminate the AGB sample.

%%%%%%%%%%%%%%%%%%%%%%%%%%%%%%%%%%%%%%%%%%%%%%%%%%%%%%%%%%%%%%%
%% Variable candidates:
\begin{deluxetable}{rrrr|rr}
\tablewidth{\columnwidth}
\tabletypesize{\normalsize}
\tablecolumns{6}
\tablecaption{Variable AGB and x-AGB Star Candidates\label{tab:vars}}

\tablehead{
\colhead{}&
\colhead{}&
\multicolumn{2}{c}{$N_{\rm AGB}$\tablenotemark{a}}&
\multicolumn{2}{c}{$N_{\rm x\mbox{-}AGB}$\tablenotemark{b}}\\
\colhead{Galaxy}&
\colhead{[Fe/H]\tablenotemark{c}}&
\colhead{($2\,\sigma$)}&
\colhead{($3\,\sigma$)}&
\colhead{($2\,\sigma$)}&
\colhead{($3\,\sigma$)}
}

\startdata
   And\,I& $-1.45 \pm 0.04$ & 0 & 1 & 0 & 3 \\          
  And\,II& $-1.64 \pm 0.04$ & 0 & 0 & 0 & 2 \\          
   And\,V& $-1.60 \pm 0.30$ & 0 & 1 & 0 & 0 \\          
  And\,IX& $-2.20 \pm 0.20$ & 0 & 0 & 1 & 4 \\          
   And\,X& $-1.93 \pm 0.11$ & 2 & 2 & 0 & 0 \\          
  And\,XI& $-2.00 \pm 0.20$ & 4 & 2 & 0 & 0 \\          
 And\,XII& $-2.10 \pm 0.20$ & 3 & 0 & 0 & 0 \\          
And\,XIII& $-1.90 \pm 0.20$ & 1 & 1 & 1 & 0 \\          
 And\,XIV& $-2.26 \pm 0.05$ & 3 & 0 & 0 & 0 \\          
And\,XVII& $-1.90 \pm 0.20$ & 0 & 0 & 1 & 0 \\          
And\,XVIII$^\dagger$& $-1.8 \pm 0.10$ &  1 & 0 & 0 & 0 \\
 And\,XIX& $-1.90 \pm 0.10$ & 1 & 0 & 0 & 0 \\          
 And\,XXI& $-1.80 \pm 0.20$ & 1 & 0 & 0 & 0 \\          
And\,XXII& $-1.62 \pm 0.05$ & 1 & 1 & 0 & 0 \\          
  Antlia$^\dagger$& $-1.60 \pm 0.10$ & 3 & 0 & 0 & 0 \\
Aquarius$^\dagger$& $-1.30 \pm 0.20$ & 3 & 0 & 0 & 2 \\ 
   Cetus& $-1.90 \pm 0.10$ & 1 & 0 & 1 & 1 \\             
 CVn\,II& $-2.20 \pm 0.05$ & 0 & 1 & 0 & 0 \\           
IC\,10$^\dagger$& $-1.28$ & 13 & 9 & 18 & 217 \\   
IC\,1613& $-1.60 \pm 0.20$ & 4 & 4 & 3 & 27 \\          
  Leo\,A& $-1.40 \pm 0.20$ & 0 & 0 & 0 & 3 \\            
  Leo\,T& $-1.99 \pm 0.05$ & 0 & 1 & 0 & 0 \\            
  LGS\,3& $-2.10 \pm 0.22$ & 1 & 1 & 0 & 1 \\            
NGC\,147$^\dagger$& $-1.10 \pm 0.10$ & 8 & 5 & 7 & 69 \\ 
NGC\,185$^\dagger$& $-1.30 \pm 0.10$ & 5 & 6 & 4 & 54 \\  
 Pegasus& $-1.40 \pm 0.20$ & 0 & 0 & 1 & 5 \\           
 Phoenix& $-1.37 \pm 0.20$ & 0 & 1 & 0 & 1\\         
Sag\,DIG$^\dagger$& $-2.10 \pm 0.20$ & 0 & 1 & 0 & 6 \\ 
Sextans\,A$^\dagger$& $-1.85$ & 1 & 0 & 1 & 23 \\   
Sextans\,B$^\dagger$& $-1.6$ & 1 & 2 & 4 & 28 \\   
  Tucana& $-1.95 \pm 0.15$ & 1 & 3 & 0 & 0 \\           
     WLM& $-1.27 \pm 0.04$ & 5 & 0 & 2 & 20              
\enddata 

\tablenotetext{a}{\ Sources are included here if they are bluer than
  the dashed line in Figure~\ref{fig:class}b and brighter than
  $M_{3.6}=-6$~mag.}

\tablenotetext{b}{\ Sources are included here if they are redder than
  the dashed line in Figure~\ref{fig:class}b.}

\tablenotetext{c}{\ Metallicities from \citet{McConnachie+2012}, also
  see Paper\,I.}

\tablenotetext{$\dagger$}{\ These galaxies have $<$75\% complete
  photometry at $M_{3.6} = -6$~mag due either to their distance or to
  crowding, though all galaxies reach 75\% completeness by
  $M_{3.6}=-6.8$~mag (Paper\,I). $N_{\rm AGB}$ should be considered a
  lower limit in these cases. Photometric incompleteness does not
  $N_{\rm x\mbox{-}AGB}$ except within the central
  $\approx$1\arcmin\ region of IC\,10 and 0.5\arcmin\ in NGC\,185.}

\tablecomments{\ Number of variable stars within the given
  color-magnitude space detected at the $2\,\sigma$ and $3\,\sigma$
  level (Section~\ref{sec:var}). This is the maximum number of
  variables allowing for the uncertainty in $(m-M)_0$ (see Table 1
  from Paper\,I). The number of sources included here are confined to
  the spatial area covered by all epochs and wavelengths (see Table~2
  from Paper\,I). }
\end{deluxetable}
%%%%%%%%%%%%%%%%%%%%%%%%%%%%%%%%%%%%%%%%%%%%%%%%%%%%%%%%%%%%%%%

%%%%%%%%%%%%%%%%%%%%%%%%%%%%%%%%%%%%%%%%%%%%%%%%%%%%%%%%%%%%%%%
\begin{deluxetable}{rrr|rr}
\tablewidth{2.5in}
\tabletypesize{\footnotesize}
\tablecolumns{5}
\tablecaption{Number of Additional AGB Star Candidates from Epoch 0\label{tab:vars3}}

\tablehead{
\colhead{}&
\multicolumn{2}{c}{$N_{\rm AGB}$}&
\multicolumn{2}{c}{$N_{\rm x\mbox{-}AGB}$}\\
\colhead{Galaxy}&
\colhead{($2\,\sigma$)}&
\colhead{($3\,\sigma$)}&
\colhead{($2\,\sigma$)}&
\colhead{($3\,\sigma$)}
}

\startdata
        Aquarius&  0&  0&  0&  0 \\
        IC\,1613&  3&  2&  1&  3 \\
          Leo\,A&  1&  0&  0&  0 \\
          LGS\,3&  1&  0&  0&  0 \\
         Pegasus&  2&  1&  0&  2 \\
         Phoenix&  0&  0&  0&  0 \\
      Sextans\,A&  0&  0&  0&  3 \\
             WLM&  0&  0&  3&  4
\enddata 

\end{deluxetable}

%%%%%%%%%%%%%%%%%%%%%%%%%%%%%%%%%%%%%%%%%%%%%%%%%%%%%%%%%%%%%%%

\subsection{Identification of Variable Sources}
\label{sec:var}

\begin{figure}
\includegraphics[width=0.85\columnwidth]{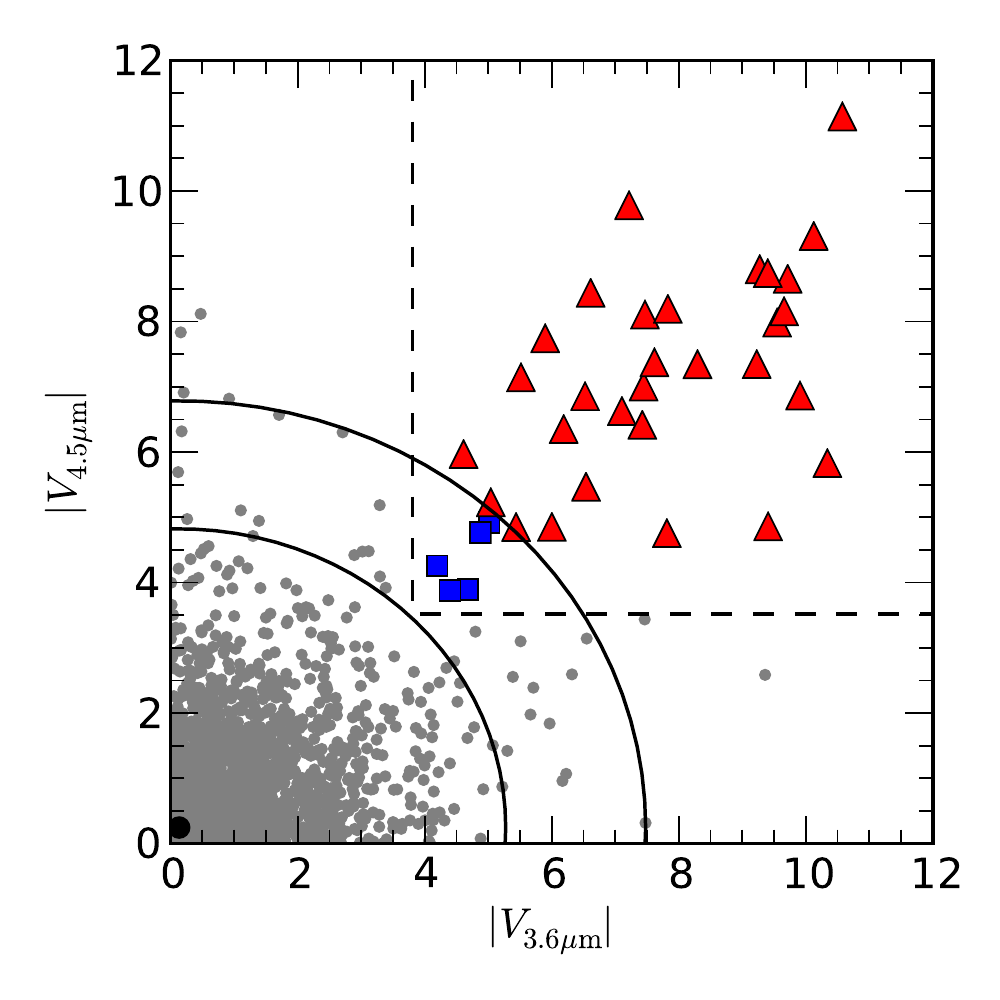}
\figcaption{Selection of variable stars in Sextans\,B.  The filled
  black circle marks the mean of the distribution. The dashed line
  marks the $2\,\sigma$ values for the absolute values of the
  variability indices at each wavelength ($\vert V_{\rm 3.6 \mu m}
  \vert$ and $\vert V_{\rm 4.5 \mu m} \vert$).  Error ellipses
  denoting the joint probability of $2\,\sigma$ and $3\,\sigma$ are
  plotted as solid lines.  Variable star candidates with high
  confidence (joint probability $>$$3\,\sigma$) are marked as filled
  red triangles.  Less confident variable star candidates (joint
  probability $>$$2\,\sigma$) are marked with filled blue
  boxes. Table~\ref{tab:vars} lists the number of variable star
  candidates selected in each target galaxy.\label{fig:vindex}}
\end{figure}

To identify individual AGB candidates, we use the variability
index defined by \citet{Vijh+09}:

\begin{equation}
V_\lambda = \frac{f_{\lambda,i} - f_{\lambda,j}}{\sqrt{\sigma f_{\lambda,i}^2 + \sigma f_{\lambda,j}^2}},
\label{eqn:var}
\end{equation}  

\noindent where $f_{\lambda}$ is the flux density of a source, $\sigma
f_\lambda$ is the uncertainty in flux, and $i$ and $j$ indicate
different epochs. Assuming a 2-dimensional Gaussian probability
distribution in $V_{\rm 3.6 \mu m}$ and $V_{\rm 4.5 \mu m}$, we
identify high-confidence variable star candidates as those with a
joint probability falling outside $3\,\sigma$. Less confident
variables stars are those falling between 2$\,\sigma$--3$\,\sigma$. In
both cases, candidates must show variability indices more than
$2\,\sigma$ outside the mean at both 3.6 and 4.5~\micron, in the same
direction. We illustrate this technique for Sextans\,B in
Figure~\ref{fig:vindex}.

For galaxies with a third epoch of data (Section~\ref{sec:cryo}), we
computed equation~\ref{eqn:var} comparing epoch 0 to each of the
DUSTiNGS epochs to identify additional variable candidates. Inclusion
of epoch 0 results in 14 more 2$\sigma$ and 19 more 3$\sigma$
variables for these 8 galaxies, including 16 variable x-AGB candidates
(Table~\ref{tab:vars3}). These additional variables are located in the
central region of each galaxy (Fig.~\ref{fig:spat1}, \ref{fig:spat2},
and \ref{fig:spat3}).

%%%%%%%%%%%%%%%%%%%%%%%%%%%%%%%%%%%%%%%%%%%%%%%%%%%%%%%%%%%%%%%
%%%%%%%%%%%%%%%%%%%%%%%%%%%%%%%%%%%%%%%%%%%%%%%%%%%%%%%%%%%%%%%

\begin{deluxetable*}{ccccccccccccc}
\tablewidth{\textwidth}
\tabletypesize{\small}
\tablecolumns{13}
\tablecaption{Variable Star Candidates\label{tab:varcandy}}

\tablehead{
\colhead{Galaxy}&
\colhead{GSC ID\tablenotemark{a}}&
\colhead{$\alpha$}&
\colhead{$\delta$}&
\colhead{$\langle m_{3.6} \rangle$}&
\colhead{$\langle m_{4.5} \rangle$}&
\colhead{$\Delta m_{3.6}$\tablenotemark{b}}&
\colhead{$\sigma_{\rm Var}$}&
\colhead{Type\tablenotemark{c}}&
\colhead{$\log \dot{D}$\tablenotemark{c}}&
\colhead{$N_{\rm epochs}$\tablenotemark{d}}&
\colhead{Flag\tablenotemark{e}}&
\colhead{Notes\tablenotemark{f}}\\
\colhead{}&
\colhead{}&
\colhead{J(2000)}&
\colhead{J(2000)}&
\colhead{(mag)}&
\colhead{(mag)}&
\colhead{(mag)}&
\colhead{}&
\colhead{}&
\colhead{$[M_\odot\,{\rm yr}^{-1}]$}&
\colhead{}&
\colhead{}&
\colhead{}
}

\startdata
  LGS\,3 &    12570  & 01 04 16.82  & $+$21 51 51.0 &  18.95 &  17.93  &  1.33 & 3 & A & \nodata & 2 & \nodata&\nodata\\
  LGS\,3 &    27414  & 01 04 05.85  & $+$21 52 39.1 &  19.20 &  18.00  &  0.94 & 2 & U & \nodata & 3 &       E&\nodata\\
  LGS\,3 &    28221  & 01 04 05.35  & $+$21 50 22.3 &  17.28 &  16.79  &  0.74 & 2 & A & \nodata & 2 & \nodata&\nodata\\
  LGS\,3 &    47308  & 01 03 55.31  & $+$21 52 47.9 &  15.22 &  14.76  &  0.37 & 3 & X & $-$8.83 & 2 & \nodata&\nodata\\
  LGS\,3 &    67343  & 01 03 43.92  & $+$21 52 30.0 &  18.54 &  17.97  &  0.64 & 2 & A & \nodata & 3 & \nodata&\nodata\\
 Phoenix &    84555  & 01 51 25.22  & $-$44 29 06.6 &  17.27 &  16.62  &  0.70 & 3 & A & \nodata & 2 & \nodata&\nodata\\
 Phoenix &    89777  & 01 51 23.01  & $-$44 29 53.6 &  18.42 &  17.91  &  1.16 & 3 & U & \nodata & 2 & \nodata&\nodata\\
 Phoenix &   102714  & 01 51 17.81  & $-$44 21 17.3 &  19.20 &  18.27  &  0.57 & 2 & U & \nodata & 2 & \nodata&\nodata\\
 Phoenix &   116894  & 01 51 12.01  & $-$44 31 58.1 &  18.84 &  18.32  &  0.72 & 2 & U & \nodata & 2 & \nodata&\nodata\\
 Phoenix &   130056  & 01 51 06.56  & $-$44 26 08.7 &  18.83 &  18.86  &  0.76 & 3 & U & \nodata & 3 & \nodata&\nodata\\
 Phoenix &   133756  & 01 51 05.06  & $-$44 22 24.8 &  17.76 &  17.93  &  0.88 & 3 & U & \nodata & 2 & \nodata&\nodata\\
 Phoenix &   143442  & 01 51 00.97  & $-$44 28 14.6 &  14.41 &  14.12  &  0.45 & 3 & X & $-$9.06 & 2 & \nodata&Mira\tablenotemark{e} \\
 Phoenix &   144095  & 01 51 00.71  & $-$44 27 54.5 &  18.61 &  18.82  &  0.48 & 3 & U & \nodata & 3 & \nodata&\nodata\\
 Phoenix &   145317  & 01 51 00.21  & $-$44 28 08.5 &  18.61 &  17.90  &  1.01 & 3 & U & \nodata & 3 & \nodata&\nodata\\
 Phoenix &   148263  & 01 50 58.99  & $-$44 23 22.5 &  19.60 &  19.00  &  0.66 & 2 & U & \nodata & 2 & \nodata&\nodata\\
 Phoenix &   206620  & 01 50 33.63  & $-$44 27 50.4 &  19.41 &  18.84  &  0.60 & 2 & U & \nodata & 2 & \nodata&\nodata

\enddata

\tablecomments{\ The variable star candidate catalog for LGS\,3 and
  Phoenix. The catalog is available for all galaxies in the electronic
  version of the paper.  The electronic catalog also includes
  magnitude and amplitude uncertainties. Stars are considered variable
  candidates if they are detected as variable at the 2 or 3\,$\sigma$
  level ($\sigma_{\rm Var}$).}

\tablenotetext{a}{\ The GSC is the ``good-source catalog'', described
  in Paper\,I.}

\tablenotetext{b}{\ We define the amplitude ($\Delta m_{3.6}$) as the
  difference between the maximum and minimum magnitude.}

\tablenotetext{c}{\ Stars are classified as described in
  Section~\ref{sec:class} as x-AGB (X), AGB (A) or unknown
  (U). Dust-production rates ($\dot{D}$) are derived only for x-AGB
  candidates, and the possible $\dot{D}$ saturation limit is not
  applied to numbers in this table (Section~\ref{sec:dpr}).}

\tablenotetext{d}{\ Most variable stars were found by comparing epochs
  1 and 2, but some were only detected as variable by including epoch
  0 (Section~\ref{sec:cryo}). For those detected via epoch 0, the mean
  magnitude in this table includes all 3 epochs.}

\tablenotetext{e}{\ Variables are flagged if they lie near a bright
  star (B), the frame edge (E) or Column Pulldown (P). The measured
  fluxes may be affected in all three cases
  (Section~\ref{sec:xagb}). We also flag stars brighter
    than the dot-dash line in Fig.~\ref{fig:class} as possible RSG
    stars.}

\tablenotetext{f}{\ If a variable candidate was identified as either a
  long-period variable or a C star in Section~\ref{sec:known}, it is
  marked in this column along with its reference. In Phoenix, star
  \#143442 is a Mira \citep[][]{Menzies+2008}.}

\end{deluxetable*}

\subsubsection{Likelihood of Detection via the Variability Index}
\label{sec:prob}

\begin{figure}
\vbox{
  \includegraphics[width=\columnwidth]{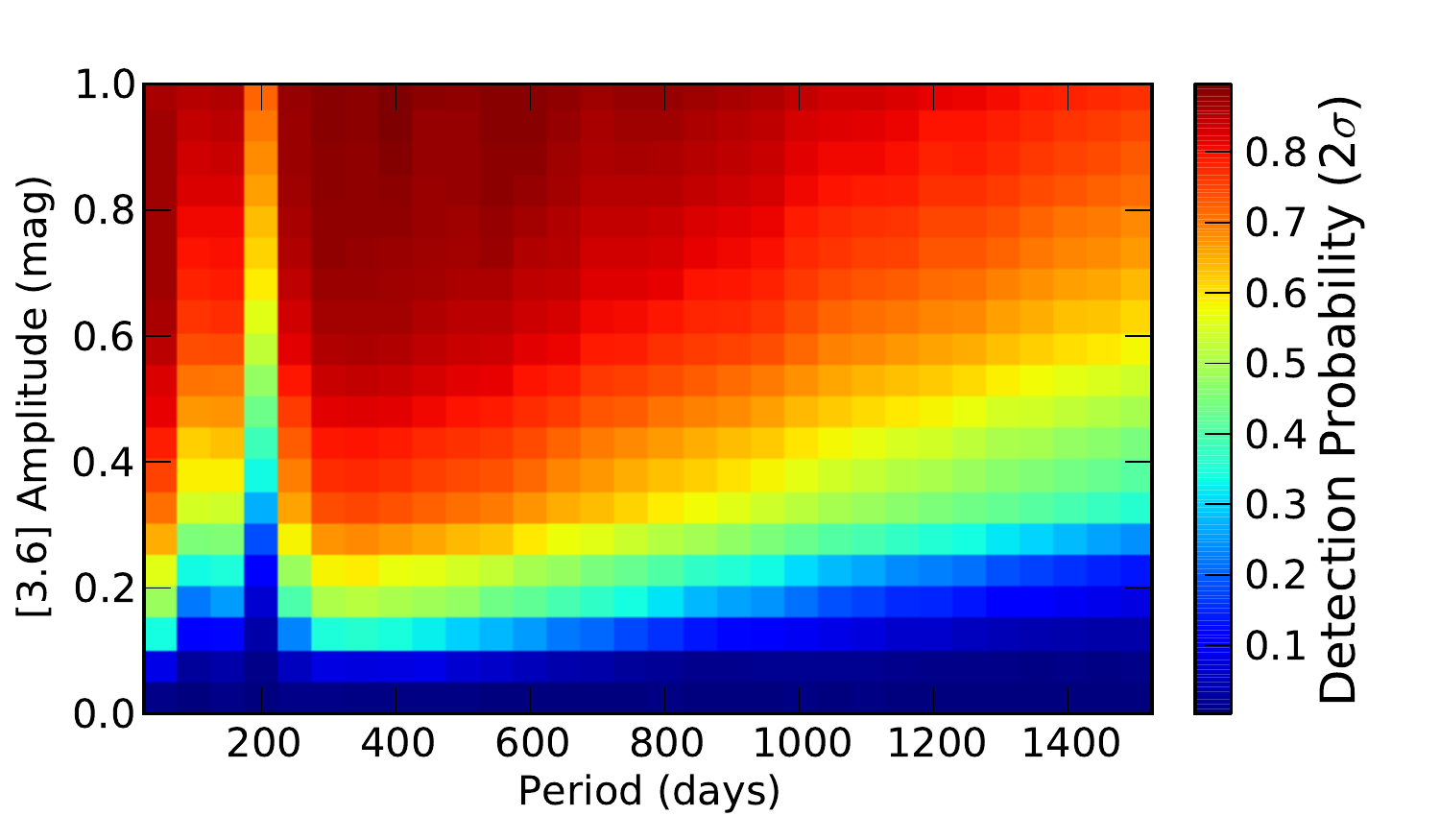}
  \includegraphics[width=\columnwidth]{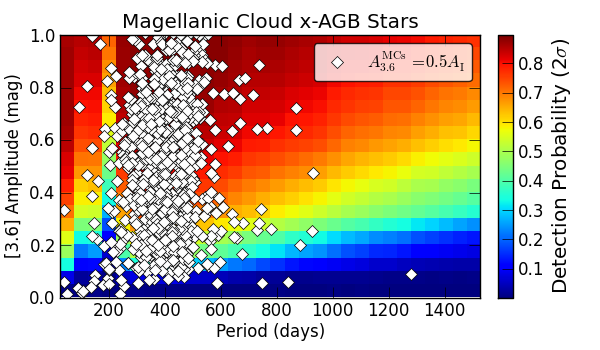}
} 

\figcaption{{\it Upper panel:} Probability of a 2\,$\sigma$ detection
  of a variable star given a separation of 180 days (6 months) between
  2 epochs. The DUSTiNGS galaxies have an epoch separation of 127--240
  days (Paper\,I). {\it Lower panel:} SMC and LMC variable x-AGB stars
  (white diamonds).  Periods and $I-$band amplitudes come from the
  OGLE survey \citep{Soszynski+08}.  We assume that the
  3.6~\micron\ amplitude is half the $I$-band
  amplitude. \label{fig:prob}}
\end{figure}

Because we have only 2--3 epochs, the cadence of the
  observations can be unfavorable for the detection of variables with
  particular amplitudes and periods. In general, the variability index
  is less sensitive to small-amplitude variable stars and stars with
  periods on the order of the separation between the epochs
  ($\approx$180~days). To determine the ``completeness'' of the
  detected variables, we compute the probabilities of detecting
  variable stars with the DUSTiNGS observations for stars with periods
  ranging from 50--3000 days and peak-to-peak amplitudes up to 1~mag
  at 3.6~\micron. Figure~\ref{fig:prob} shows the result of this
simulation. For comparison, we also show the SMC and LMC
optically-detected variable x-AGB stars \citep[from the Optical
  Gravitational Lensing Experiment III,
  OGLE\,III;][]{OGLE3_LMC,OGLE3_SMC}.  The DUSTiNGS variability index
is insensitive to stars with amplitudes less than $\approx$0.15~mag
(especially less-evolved O-rich AGB stars; Fig.~\ref{Afig:sim}), but
it is particularly suited to detect the x-AGB stars with periods
ranging from 300--600~days.

This simulation assumes that the 3.6~\micron\ amplitude is equal to
half the $I$-band amplitude, since it is known that the amplitude
tends to decrease at near-IR wavelengths
\citep[e.g.,][]{Whitelock+2006,Whitelock+2009,Menzies+2008}. Little is
known about typical AGB star amplitudes at IRAC wavelengths; for
IRC$+$10216, the $L^\prime$ (3.79~\micron) amplitude is $\approx 0.75
A_{\rm J}$ \citep{LeBertre1992}. The most extreme, optically-obscured
AGB stars ($[3.6]-[4.5] \gtrsim 0.9$~mag) may be more easily detected
by DUSTiNGS if they follow the general trend for more evolved stars
having larger amplitudes \citep[e.g., the C star LI-LMC
  1813;][]{vanLoon+2003}.

%%%%%%%%%%%%%%%%%%%%%%%%%%%%%%%%%%%%%%%%%%%%%%%%%%%%%%%%%%%%%%%
%%%%%%%%%%%%%%%%%%%%%%%%%%%%%%%%%%%%%%%%%%%%%%%%%%%%%%%%%%%%%%%

\subsubsection{Contamination Among Variable Candidates}
\label{sec:contam}

With 2 epochs of IRAC data, spaced 3~months apart, \citet{Vijh+09}
conducted a similar variability analysis for the LMC with data from
the SAGE program \citep{Meixner+06}. Using additional optical and
near-IR photometry, they classified sources into different types of
AGB stars (carbon-rich, oxygen-rich, and x-AGB stars), using the
classification scheme outlined by \citet{Blum+06}.  They found that
the most (92\%) of the variable stars detected in the LMC
above the TRGB are one of these AGB types. young stellar objects
(YSOs), as classified by \citet{Whitney+08}, comprise 15\% of the
remaining variable sources.  The rest were unclassified, and may
include Cepheids, RR~Lyrae stars, and eclipsing binaries. Based on
their results, we expect non-AGB variable stars to comprise only a
small percentage of the variable candidates detected in the
DUSTiNGS galaxies (see Fig.~\ref{Afig:sim}a).

Since x-AGB stars are more evolved than regular O-rich and C-rich AGB
stars, their periods are longer and their amplitudes are larger,
making them the easiest stars to detect with only 2 epochs. Almost
half of the variable stars detected by \citet{Vijh+09} in the LMC
above the TRGB are x-AGB stars, despite the fact that these very
dusty, short-lived stars comprise only 4.5\% of the total LMC AGB
population \citep{Boyer+11}.

It is also possible that variable sources are background AGN, which
are known to vary irregularly \citep[for a review,
  see][]{Ulrich+2007}. Most of the known background AGN are fainter
than 16~mag at 3.6~\micron\ \citep[e.g.,][]{Sanders+2007}, so we do
not expect much AGN contamination among the x-AGB variable stars. In
addition, even though we observe a large population of faint and red
sources, few red sources fainter than the x-AGB candidates are
detected as variable here. This suggests that the DUSTiNGS survey is
not sensitive to AGN amplitudes and/or cadences.

\section{Results}
\label{sec:results}

The results of the color and variable classifications
  are listed in Tables~\ref{tab:vars}, \ref{tab:vars3}, and
  \ref{tab:varcandy}. In this section, we describe the observed properties of the x-AGB variable stars.

\subsection{Variable x-AGB Star Candidates}
\label{sec:xagb}

%%%%%%%%%%%%%%%%%%%%%%%%%%%%%%%%%%%%%%%%%%%%%%%%%%%%%%%%%%%%%%%
\begin{figure}
\vbox{
  \includegraphics[width=\columnwidth]{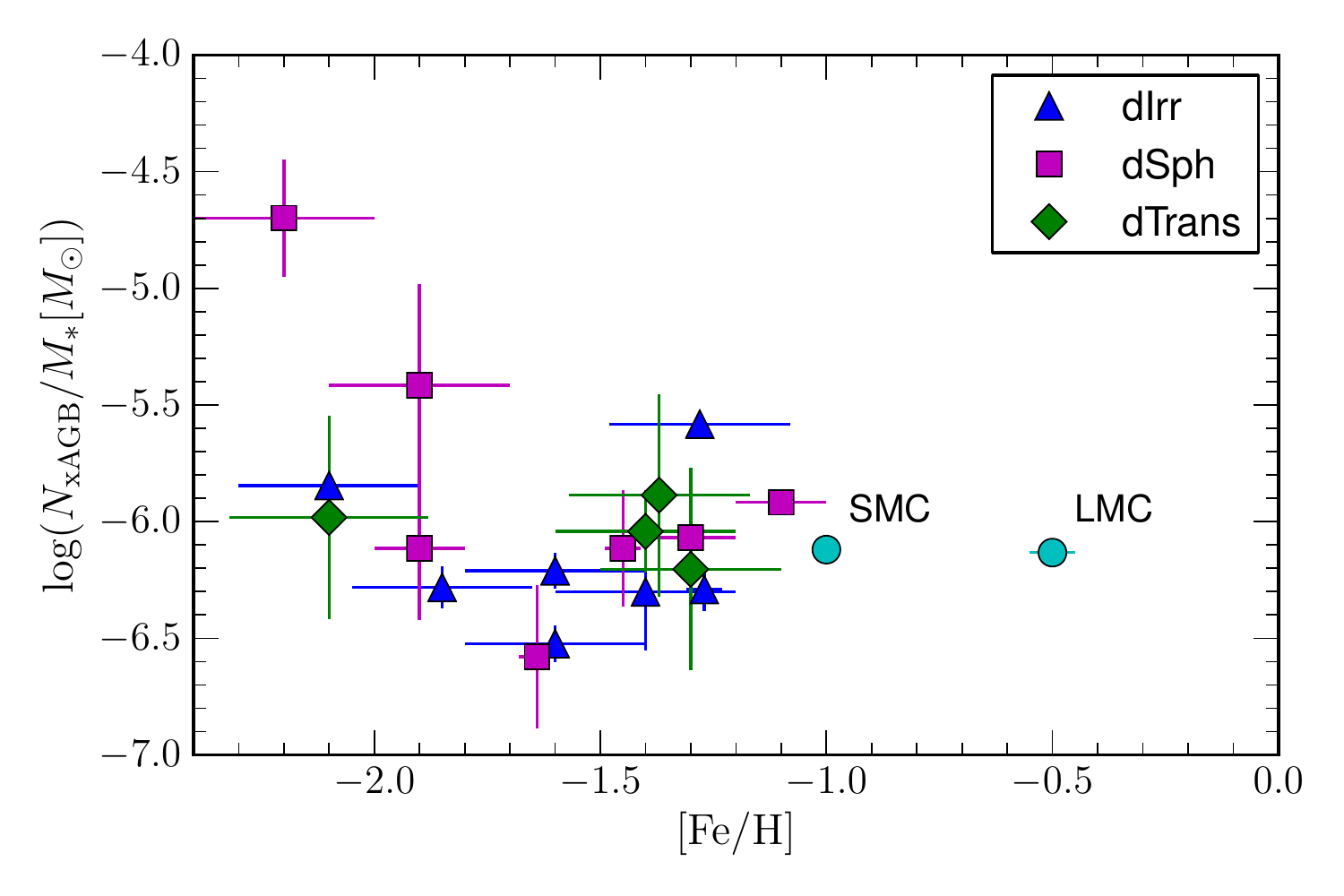}
}
\caption{The number of x-AGB variable candidates normalized to the
  total stellar mass as a function of metallicity. Masses and
  metallicities are from \citet{McConnachie+2012} and are listed in
  Paper\,I. Y-axis error bars reflect Poisson uncertainties only. The
  square in the upper left is And\,IX, but imaging artifacts may
  affect 3 of the 5 x-AGB variables in this galaxy (see
  Section~\ref{sec:xagb}). This plot includes only variables detected
  from epochs 1 and 2 to avoid artificial inflation of the number of
  variables detected in the subset of galaxies with epoch 0
  data.\label{fig:nxagb}}
\end{figure}
%%%%%%%%%%%%%%%%%%%%%%%%%%%%%%%%%%%%%%%%%%%%%%%%%%%%%%%%%%%%%%%

In the entire DUSTiNGS sample, we find 526 variable
  x-AGB candidates among 710 total variable sources. The number of
  variable AGB star candidates in each galaxy is listed in
  Tables~\ref{tab:vars} and \ref{tab:vars3}. We quote the maximum
  number of AGB candidates allowing for the uncertainty in the
  distance moduli (Paper\,I).  Individual variable sources are listed
  in Table~\ref{tab:varcandy} along with mean magnitudes, the lower
  limits on the 3.6~\micron\ amplitude ($\Delta m_{3.6}$ = $\vert
  m_{3.6}^{\rm epoch\,1} - m_{3.6}^{\rm epoch\,2} \vert$), and the
  classification. This table is available to download in the
  electronic version of this paper.

We have flagged a small subset of the variable sources listed in
Table~\ref{tab:varcandy} that may be affected by imaging artifacts
that cause the star to appear artificially variable.  Some x-AGB
candidates are located near imaging artifacts caused by ``Column
Pulldown''\footnote{\label{irac}http://irsa.ipac.caltech.edu/data/SPITZER/docs/irac/\\iracinstrumenthandbook/home/},
which can be seen as lines of faint pixels emanating from bright point
sources.  This affects 11 variable x-AGB star candidates in And\,IX,
And\,XIII, Aquarius, IC\,10, NGC\,147, Sag\,DIG, and Sextans\,A.  In
addition, some variable sources fall on the edge of the spatial
coverage ($\lesssim$10\arcsec\ from the edge) where fewer frames
contribute to the total depth owing to frame dithering (Paper\,I).
There are fewer frames to assist in the elimination of imaging
artifacts in these regions, which may affect the measured fluxes.
This affects 4 variable x-AGB candidates, including 2 in And\,IX, 1 in
IC\,1613, and 1 in IC\,10. We also checked whether proximity to a
bright star might affect the measured flux and found 1 variable x-AGB
star in NGC\,147 that may be artificially variable. We exclude
affected sources from further analysis, with the exception of source
\#21181 in And\,IX at $\alpha$(J2000) = $13^{\rm h}18^{\rm m}54\fs04$,
$\delta$(J2000) = $+43\degr12\arcmin2\farcs8$. This star may be
marginally affected by Column Pulldown at 3.6~\micron\ in epoch 2 and
4.5~\micron\ in epoch 1, but since the color is similar between epochs
and the amplitude at each wavelength is similar, we include it in the
analysis as a high-confidence x-AGB star.

In Paper\,I, we estimated the total size of the x-AGB population by
statistically subtracting the foreground and background sources. For
the 6 galaxies with large x-AGB populations and firm (non-upper-limit)
estimates from Paper\,I, the fraction of x-AGB stars detected as
variable ranges from 33\%--61\% (this includes IC\,10, IC\,1613,
NGC\,147, NGC\,185, Sextans\,B, and WLM). Using the AGB star catalogs
for the LMC and SMC from \citet{Boyer+11} and the OGLE $I$-band
periods, we find that a DUSTiNGS-like survey would detect
$\approx$62\% of the OGLE-detected MC x-AGB stars (probability of a
2-$\sigma$ detection $>$0.75 with 1000 trials;
Section~\ref{sec:prob}). The difference in detection fractions is more
likely a reflection of the differences in the distance to a galaxy and
the epoch separation, and not to a real difference in the pulsation
amplitudes between galaxies. A more distant galaxy has larger
photometric uncertainties resulting in a lower probability of
detection, and a longer epoch separation results in a higher
probability of detection, in general.

Figure~\ref{fig:nxagb} shows the total number of x-AGB variable
candidates, normalized to each galaxy's total stellar mass, as a
function of metallicity, using the masses and metallicities from
\citet{McConnachie+2012}.  There is no clear metallicity trend for the
number of x-AGB stars in a galaxy, suggesting that the dominant factor
determining the number of x-AGB stars is the star formation history.

\subsection{Color-Magnitude Diagram}
\label{sec:cmd}

\begin{figure}
\includegraphics[width=\columnwidth]{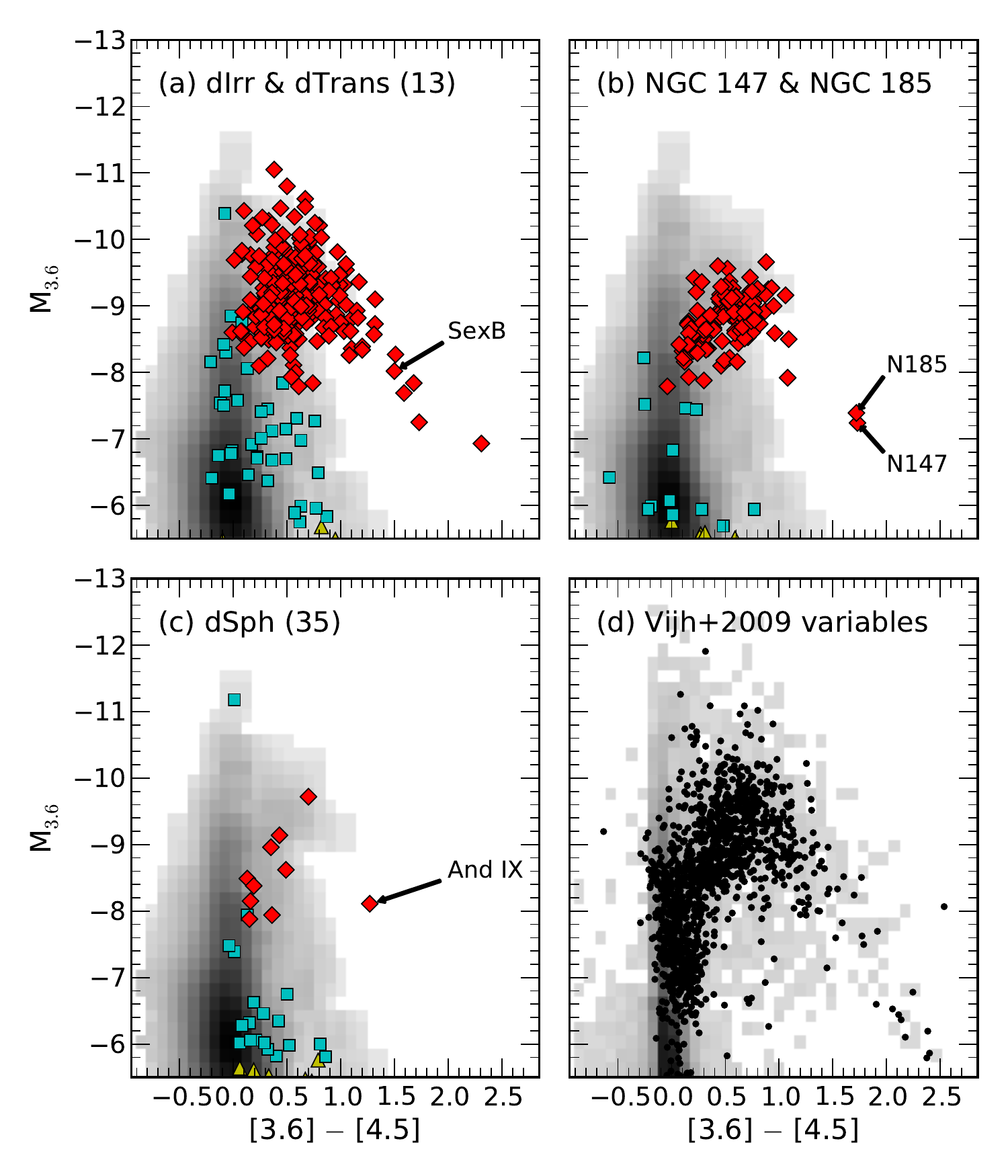}
\caption{Location of variable sources on the CMD.  The x-AGB variable
  star candidates are plotted as red diamonds, AGB variable star
  candidates as cyan squares, and unknown variables as yellow
  triangles. The greyscale background in all panels is the mean
  background-subtracted CMD (plotted as the density of sources) for
  all 50 target galaxies. {\it (a):} Variable sources from the 13 dIrr
  and dTrans galaxies.  All x-AGB stars redder than $[3.6]-[4.5] =
  1.5$~mag are in IC\,10, except the one star labeled Sextans\,B. {\it
    (b):} Variable sources from NGC\,147 and NGC\,185, which we plot
  separately since they dominate the dSph x-AGB populations. {\it
    (c):} Variable sources from the remaining 35 dSph galaxies.  {\it
    (d):} Variable sources detected with 2 epochs separated by 3
  months in the LMC \citep{Vijh+09}, plotted over the same greyscale
  background shown in Figure~\ref{fig:class}a. \label{fig:varcmd}}
\end{figure}

In Paper\,I, we showed that the CMDs for most of the
  DUSTiNGS galaxies are dominated by foreground and/or background
  sources, making it difficult to see the true color distribution of
  AGB stars. In Figure~\ref{fig:varcmd}, we show the location of the
variable candidates overplotted on the CMD, separated by the
morphological type of the galaxy.  The greyscale in panels {\it a--c}
is the average background-subtracted CMD for all 50 galaxies. For
comparison, Figure~\ref{fig:varcmd}d includes a similar diagram with
the variables identified in the LMC by \citet[][see
  Section~\ref{sec:contam}]{Vijh+09}. We detect few blue AGB stars
compared to the LMC, though x-AGB stars are recovered at a similar
rate.  This is due to dustless stars pulsating with amplitudes that
are smaller than the DUSTiNGS photometric uncertainty
(Fig.~\ref{Afig:sim}).

Most of the x-AGB star candidates with $[3.6]-[4.5] \gtrsim 1.5$~mag
are in IC\,10, which by far harbors the largest x-AGB population
(Table~\ref{tab:vars}) and is thus the most likely to be host to rarer
stellar types. Otherwise, only one star each from Sextans\,B,
NGC\,185, and NGC\,147 is as red as the IC\,10 x-AGB stars
(Fig.~\ref{fig:varcmd}). With the exception of NGC\,147 and NGC\,185,
the dSph galaxies harbor few examples of x-AGB candidates, as
expected based on their star formation histories.  Altogether, we find
only 10 x-AGB star candidates in the remaining 35 dSph galaxies. Star
\#21181 in And\,IX is particularly red (Fig.~\ref{fig:varcmd}c),
indicating strong dust production (Section~\ref{sec:dpr}).

The variable x-AGB candidates are most likely C-rich AGB stars, as is
indicated in the MCs \citep[e.g.,][]{Woods+2011,Riebel+2012}.
However, a small fraction might be O-rich, especially the brightest
examples near the RSG border in Figure~\ref{fig:class}a where
hot-bottom burning might disrupt the dredge-up of carbon.  It is
impossible to distinguish the different chemistries without near-IR
photometry or spectra (Section~\ref{sec:cstars}).

\subsection{Spatial Distributions}
\label{sec:spat}

Most of the x-AGB variable stars are confined to within 1--2
half-light radii ($r_{\rm h}$;
Figs.~\ref{fig:spat1}--\ref{fig:spat3}). However, in some galaxies the
x-AGB stars can be found at large radii (e.g., IC\,10).  In
Sextans\,B, x-AGB stars fill the full DUSTiNGS coverage
($\approx$$13\arcmin \times 14\arcmin$). This is much more extended
than the small half-light radius listed in \citet{McConnachie+2012}
($r_{\rm h} = 0\farcm9$), and agrees better with the new estimate from
\citet{Bellazzini+2014} of $r_{\rm h} = 1\farcm9$.

As in the LMC and SMC \citep{Blum+06,Sandstrom+09,Boyer+11}, the x-AGB
star distributions are generally smooth and symmetric, with a few
exceptions. In Sextans\,B, the x-AGB stars beyond the half-light
radius are preferentially located to the north, which may be due to
the influence of Sextans A and NGC 3109, which are located nearby and
towards the south. In Sextans\,A, x-AGB stars are to the southwest of
the center, avoiding the region of star formation on the east side of
the galaxy. NGC\,185 is lacking x-AGB stars to its north and south,
despite the almost circular distribution of other stellar types in the
galaxy. In IC\,10, the x-AGB candidates avoid a region of strong star
formation just to the southeast of the galaxy's center, which may be
due in part to crowding (Paper\,I).  Finally, in NGC\,147 there is a
lack of AGB stars on the southeast side of the galaxy. A detailed
examination of the radial distribution of x-AGB stars in each galaxy
will be presented in a forthcoming paper.

%%%%%%%%%%%%%%%%%%%%%%%%%%%%%%%%%%%%%%%%%%%%%%%%%%%%%%%%%%%%%%%
%%%%%%%%%%%%%%%%%%%%%%%%%%%%%%%%%%%%%%%%%%%%%%%%%%%%%%%%%%%%%%%

\subsection{IR Amplitudes}
\label{sec:iramps}

In general, pulsation amplitudes increase with periods
\citep[e.g.,][]{VassiliadisWood93,Whitelock+2000}. Thus, amplitudes
will increase as a star evolves, and larger amplitudes may thus be
linked with more dust formation.  With only two or three epochs, we
can present only a lower-limit amplitude for each variable star
($\Delta m = \vert m^{\rm epoch\,1} - m^{\rm epoch\,2} \vert$;
Table~\ref{tab:varcandy}). The mean $\Delta m_{3.6}$ detected for
x-AGB stars is $0.50 \pm 0.23$~mag (standard deviation).  Using the
variable catalog from \citet{Vijh+09} in the LMC, we compute that the
same $\Delta m_{3.6}$ for 820 x-AGB stars in the LMC is $0.36 \pm
0.19$~mag. The lower LMC mean amplitude is probably due to the smaller
distance to the LMC. As a result, photometric uncertainties in the LMC
data are smaller and enable the detection of smaller amplitudes.

The maximum amplitude is for a star in Sextans\,B ($\Delta m_{3.6} =
1.6$~mag) that has a mean color of $[3.6]-[4.5]=0.61$~mag and is
located on the outskirts of the galaxy (upper right of
Fig.~\ref{fig:spat2} at $\alpha$(J2000)$ =09^{\rm h}59^{\rm
  m}46\fs07$, $\delta$(J2000)$ =+05\degr26\arcmin08\farcs8$). The
amplitude of this star is similar to the largest $\Delta m_{3.6}$
detected for an x-AGB star in the LMC ($\Delta m_{3.6}^{\rm max} =
1.8$~mag). Since amplitudes tend to decrease with wavelength for
long-period variables \citep[e.g.,][]{LeBertre1992}, the large
amplitudes detected in some DUSTiNGS stars may imply amplitudes of
several magnitudes in the optical.  An alternate explanation is that
the large IR amplitude is the result of a varying temperature of warm
circumstellar dust. Since 3.6~\micron\ may be measuring the Wien tail
of the dust blackbody, even small changes in the total dust opacity or
temperature could result in large changes at 3.6~\micron. Changes
in the strength of molecular absorption features (possibly from dust
veiling) might also account for amplitude changes with wavelength.

\begin{figure}
\vbox{
\includegraphics[width=\columnwidth]{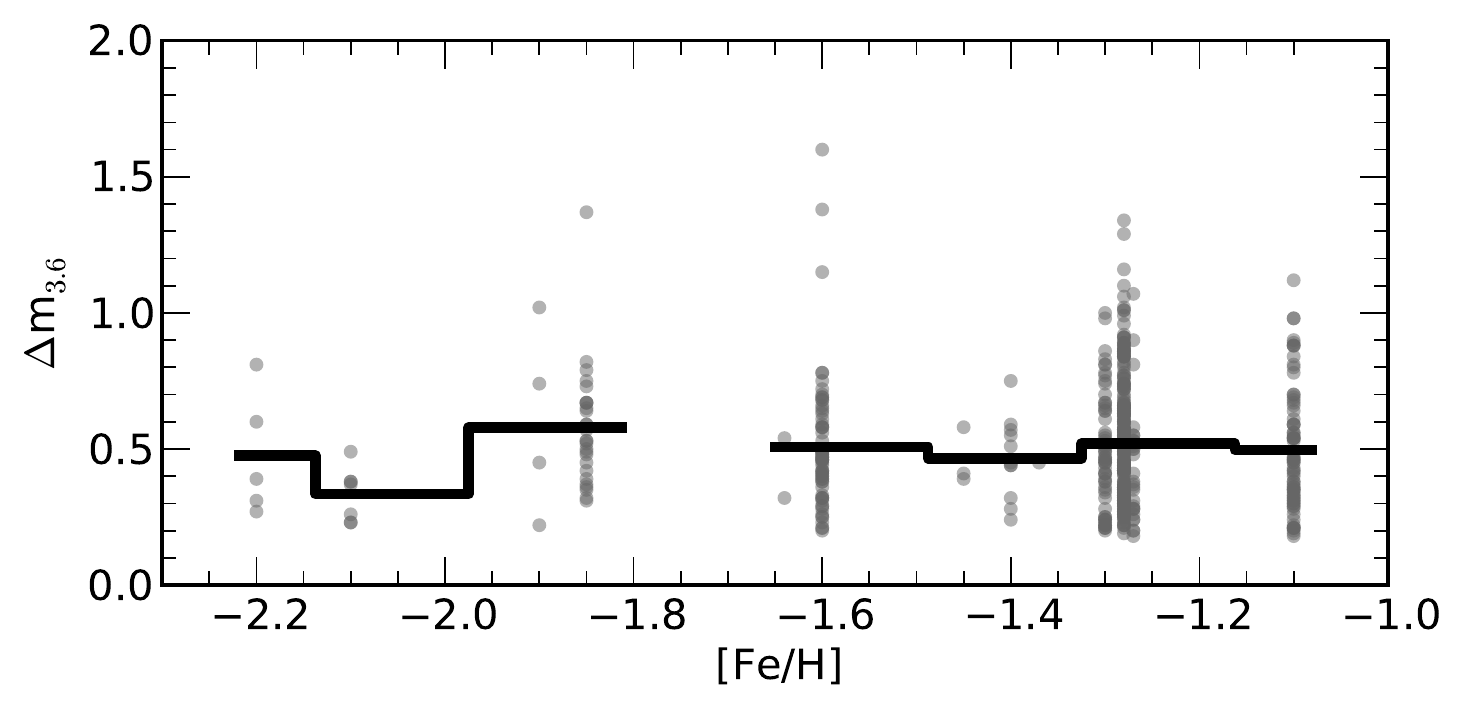}
\includegraphics[width=\columnwidth]{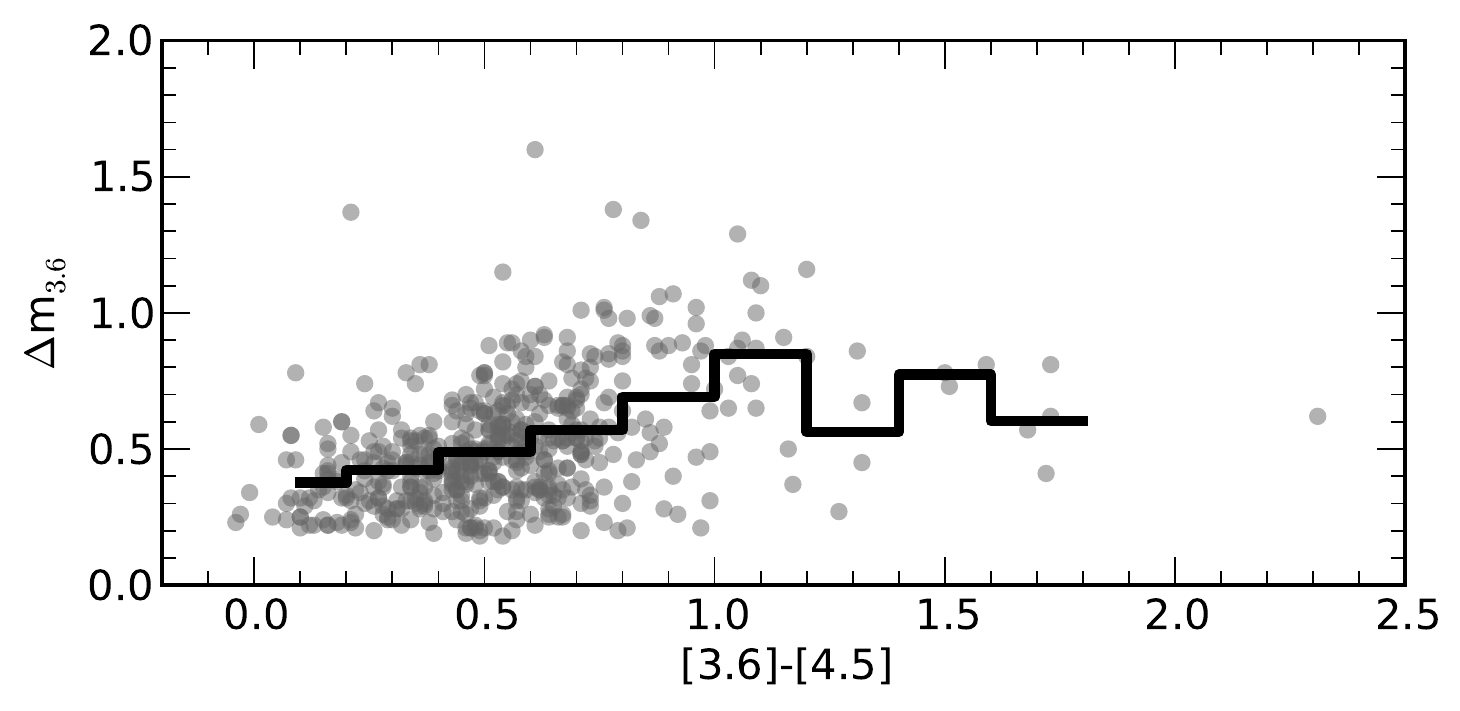}
}
\caption{IR amplitudes of x-AGB variables. Dots in these diagrams are
  transparent; regions that appear darker gray thus indicate a higher
  density of points. {\it Top Panel:} The minimum amplitude as a
  function of metallicity. The histogram marks the mean of the plotted
  points within the given metallicity bin. The mean 2-epoch minimum
  amplitude of LMC x-AGB stars from \citet{Vijh+09} is 0.36~mag at
  3.6~\micron. {\it Bottom Panel:} Amplitude as a function of
  $[3.6]-[4.5]$, with the mean amplitude within color bins marked by
  the solid line. Even with only 2 epochs, a trend for increasing
  color with amplitude is clear. \citet{McQuinn+2007} see a similar
  trend in M33 at the same wavelengths.  There is no strong trend
  linking color and metallicity (see
  Fig.~\ref{fig:varcmd_byZ}). \label{fig:amp}}
\end{figure}

In Figure~\ref{fig:amp}, we show $\Delta m_{3.6}$ for each variable
x-AGB star as a function of the host galaxy's metallicity. There is no
clear trend indicating a change in pulsation properties in metal-poor
environments.

Figure~\ref{fig:amp} also shows $\Delta m_{3.6}$ as a function of
color.  Even with only 2 epochs, it is evident that the amplitude
tends to increase with $[3.6]-[4.5]$.  This trend supports a direct
link between the star's pulsation and the dust production, and adds to
evidence of the same link seen in the Galaxy, M33, and Sgr dSph
variable AGB stars \citep{Whitelock+2006,McQuinn+2007,McDonald+2014}.

The difference between the 3.6 and 4.5~\micron\ amplitudes of x-AGB
stars is small, with $\langle\Delta m_{3.6} - \Delta m_{4.5}\rangle =
0.02 \pm 0.11$~mag. There is no trend for a systematic increase or
decrease in the amplitudes from 3.6 to 4.5~\micron.

%%%%%%%%%%%%%%%%%%%%%%%%%%%%%%%%%%%%%%%%%%%%%%%%%%%%%%%%%%%%%%%
%%%%%%%%%%%%%%%%%%%%%%%%%%%%%%%%%%%%%%%%%%%%%%%%%%%%%%%%%%%%%%%

\section{Previously Detected AGB Stars}
\label{sec:known}

The red colors and variability of DUSTiNGS x-AGB candidates support
the strong likelihood that they are true dust-producing AGB
stars. However, with only 2 epochs of 3.6 and 4.5~\micron\ imaging,
they remain candidates. In this section, we cross-identify the
DUSTiNGS sources with existing optical and near-IR surveys to confirm
some of the x-AGB candidates as C stars and/or long-period variables
(LPVs).

\subsection{Comparison to Known Variables}
\label{sec:knownvar}

All TP-AGB stars are LPVs ($20 \lesssim P \lesssim 1000$~days),
pulsating regularly or semi-regularly during different phases of their
evolution \citep[e.g.,][]{Fraser+08,Soszynski+2009}. At the end of their evolution, AGB stars become Miras,
which pulsate in the fundamental mode and generally have red colors,
large amplitudes, and periods longer than 100~days
\citep[][]{IbenRenzini1983}. Miras and other LPVs have previously been
detected in only 4 of the DUSTiNGS galaxies: NGC\,147, NGC\,185,
IC\,1613, and Phoenix.

\subsubsection{NGC\,147 and NGC\,185}
\label{sec:ngc}

\citet{Lorenz+2011} (hereafter, L$+$11) searched for LPVs in NGC\,147
and NGC\,185 with 30 epochs of photometry in the $i$-band. They found
182 and 387 LPVs, respectively, that are also detected in the $K_{\rm
  S}$-band.  Their survey includes stars within $R\sim3.5\arcmin$, or
about half the radius probed with DUSTiNGS. We find only 6 variables
in common with NGC\,147 and 22 with NGC\,185. The $i$-band amplitudes
measured by L$+$11 are $\approx$0.2--2~mag. Since we have detected
only a small fraction of the L$+$11 LPVs, it follows that the
amplitudes of the stars covered by L$+$11 are significantly smaller at
3.6~\micron\ than in the $i$-band, on average. For the variable stars
detected both here and by L$+$11, the mean amplitudes are 0.5~mag
larger in the $i$-band than at 3.6~\micron\ ($\langle A_i \rangle =
0.9$~mag and $\langle A_{3.6} \rangle = 0.4$~mag).

\citet{Nowotny+2003} used narrow-band photometry centered on the TiO
and CN bands to distinguish O-rich from C-rich AGB stars in NGC\,147
and NGC\,185 (also see Section~\ref{sec:cstars}). In NGC\,185, the 24
L$+$11 LPVs detected as variable here include 5 O-rich stars (M type),
5 C-rich stars, and 4 stars with ${\rm C/O} \approx 1$ (S type).  Most
of the M stars that are also LPVs in L$+$11 are within 2~mag of the
assumed TRGB, but of the 5 M stars that are detected as variable here,
4 are bright and red enough to be classified as x-AGB variables. This
suggests that these 4 stars may be massive enough to undergo
hot-bottom burning. Similarly, in NGC\,147, we detect 2 M type, 3
C-rich, and 1 S type star. One of these M-type LPVs is quite bright at
$M_{3.6}=-9.7$~mag and may therefore be a massive AGB star or a red
supergiant.

\begin{figure}
\includegraphics[width=\columnwidth]{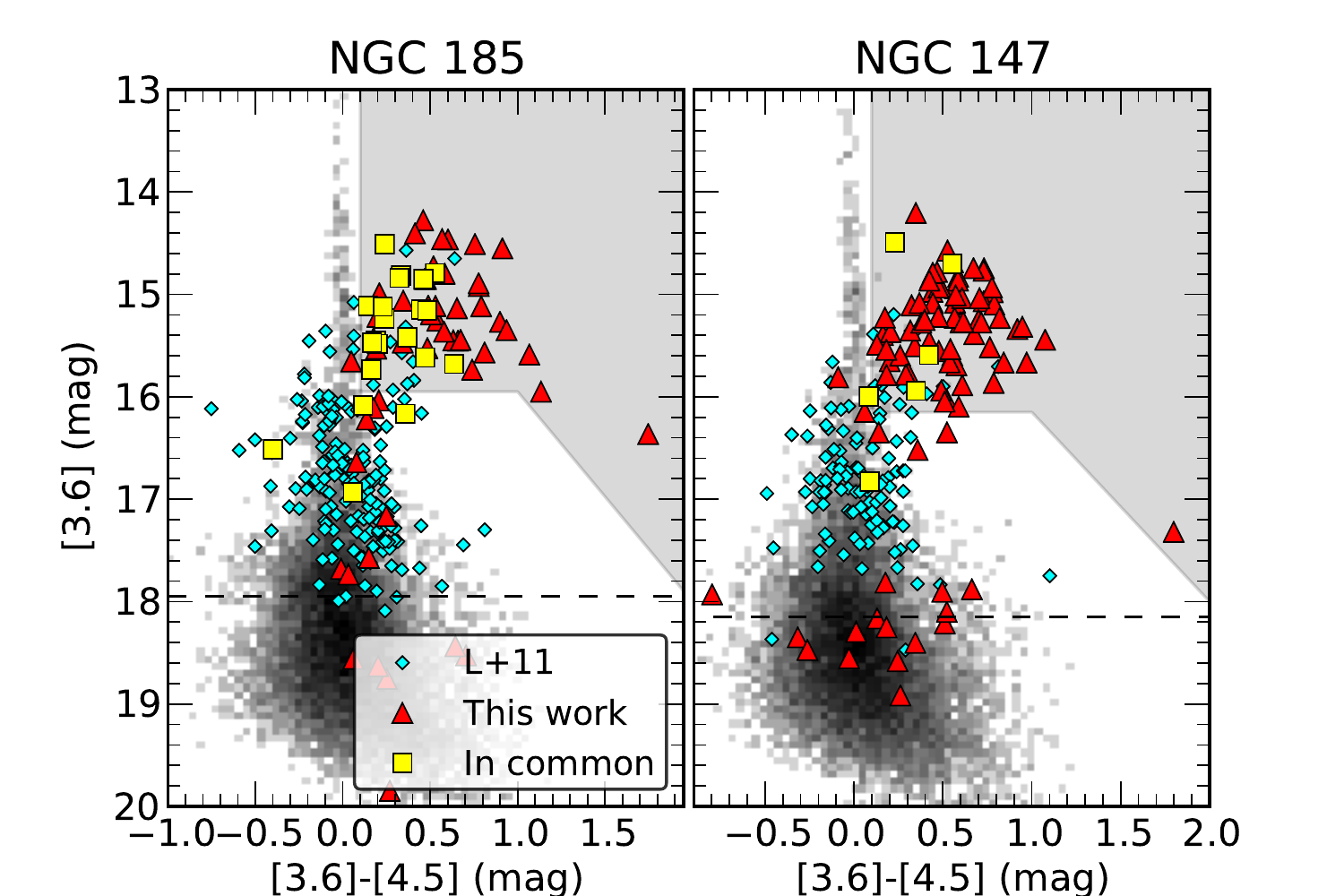}
\caption{LPVs from \citet{Lorenz+2011}. Variable sources in common
  between this work and L$+$11 are marked with yellow squares. The
  shaded area marks the region used to identify x-AGB candidates and
  the dashed line marks the assumed location of the
  TRGB. \label{fig:ngc}}
\end{figure}

Figure~\ref{fig:ngc} shows the DUSTiNGS CMD for these galaxies with
L$+$11 variables marked. It is clear that optical and near-IR surveys
are biased against the reddest sources that are uncovered by DUSTiNGS.
This also suggests that red LPVs have larger pulsation amplitudes at
3.6~\micron\ than bluer LPVs (Section~\ref{sec:iramps}). Within the
same coverage surveyed by L$+$11, we detect 38 and 28 new variable
star candidates in NGC\,147 and NGC\,185.  Of these, most are x-AGB
candidates.

%%%%%%%%%%%%%%%%%%%%%%%%%%%%%%%%%%%%%%%%%%%%%%%%%%%%%%%%%%%%%%%
%%%%%%%%%%%%%%%%%%%%%%%%%%%%%%%%%%%%%%%%%%%%%%%%%%%%%%%%%%%%%%%

\subsubsection{IC\,1613}

\citet{Kurtev+2001} (hereafter K$+$01) detected one
M-type Mira in IC\,1613 at $\alpha$(J2000) = $1^{\rm h}04^{\rm
  m}48\fs97$, $\delta$(J2000) = $+2\degr05\arcmin28\farcs8$.  This
star is also detected here as a 3\,$\sigma$ variable with $\langle
m_{3.6} \rangle= 14.63$~mag and $\langle [3.6]-[4.5] \rangle =
0.16$~mag, which qualifies it as an x-AGB candidate. K$+$01 measured a
period of 641~d and an $R$-band amplitude $>$2.5~mag. The lower limit
on the 3.6~\micron\ amplitude measured here is $\Delta m_{3.6} =
0.43$~mag.

Several works \citep[most recently,][]{Antonello+2000,
  Mantegazza+2001, Dolphin+2001} have searched for variable stars in
IC\,1613, focusing on short-period Cepheids and RR\,Lyr stars. K$+$01
is the first to report the detection of a Mira.  Here, we detect 50
new variable AGB candidates, of which 34 are x-AGB candidates.

\subsubsection{Phoenix}

\citet{Menzies+2008} reported a single Mira in Phoenix (their star \#51)
with a period of 425~d and an amplitude of $\Delta K_{\rm S} =
0.76$~mag. Its spectral type is unknown, though \citet{Menzies+2008}
argued it is C-rich based on its $J-K_{\rm S}$ color.  Here, this star
is a 3\,$\sigma$ x-AGB variable candidate with $\langle m_{3.6}
\rangle = 14.41$~mag and $\langle [3.6]-[4.5] \rangle = 0.29$~mag. Its
lower-limit amplitude is $\Delta m_{3.6} = 0.45$~mag.

\citet{Gallart+2004} identified 5 additional LPV candidates.
\citet{Menzies+2008} studied two of these and found that they are not
variable. One of these 5 is not detected in DUSTiNGS despite falling
inside the DUSTiNGS spatial coverage. Of the remaining 4 LPV
candidates, none are detected as variable here, though their colors
and magnitudes suggest that they are possible AGB stars.  One ($1^{\rm
  h}50^{\rm m}47\fs07$, $-44\degr27\arcmin43\farcs0$) shows
$[3.6]-[4.5]=1.0$~mag and $M_{3.6}=-7.28$~mag, suggesting that it
could be a dusty AGB star.

\subsection{Comparison to Known Carbon Stars in the x-AGB Sample}
\label{sec:cstars}

Carbon stars form more easily at low metallicity since less free
oxygen is available to tie up newly formed carbon into CO
molecules. Analysis of the IR spectra and spectral energy distributions
(SEDs) has revealed that most of the x-AGB stars in the LMC
and SMC are C-rich \citep[][Ruffle et al., in
  preparation]{Woods+2011}, and it follows that C stars produce
the majority of the AGB dust budget. Several of the DUSTiNGS galaxies have
been observed using optical narrow-band CN/TiO photometry or $JHK$
photometry to identify C stars.  The near-IR $JHK$ photometry is
less precise than CN/TiO photometry, though at subsolar metallicities,
the stars with red $J-K$ colors can be classified as high-confidence C
stars \citep{Cioni+06b,Aringer+2009,Boyer+2013}.

Neither method is sensitive to the dustiest stars because of
circumstellar extinction at optical and near-IR wavelengths. Optical
photometry can detect stars as red as $[3.6]-[4.5] \approx 0.5$~mag,
while the reddest DUSTiNGS star detected via $J-K$ color has
$[3.6]-[4.5]=0.9$~mag (a star in Sag\,DIG, see below). Altogether, we
confirm 70 variable x-AGB candidates as C stars. These stars are
flagged in Table~\ref{tab:varcandy}.

\subsubsection{NGC\,185}

\citet{Nowotny+2003} (hereafter N$+$03) obtained narrow-band CN/TiO
photometry of NGC\,185 and find 154 C stars. We detect 93 of the
N$+$03 C stars, which have a mean color $\langle [3.6]-[4.5] \rangle =
0.05$~mag. Of these 93, 22 would be classified as x-AGB candidates
based on their DUSTiNGS photometry ($([3.6]-[4.5])_{\rm
  max}=0.53$~mag).

Among the DUSTiNGS variable x-AGB stars, N$+$03 classified 11 as
C-rich, 3 as O-rich, and 3 as ``other'', where ``other'' indicates
that the star did not show strong CN or TiO. Similarly, 2 DUSTiNGS
variable AGB stars are classified as C-rich, and one as ``other''.

\citet{BattinelliDemers2004b} also used narrow-band CN/TiO optical
filters to identify C stars in NGC\,185, though they choose a slightly
different CN/TiO index cut-off to classify the stars. From their
C star catalog, we find 3 additional variable x-AGB candidates that are
classified as C-rich for a total of 14 C-rich variable x-AGB stars in
NGC\,185.

\subsubsection{NGC\,147}

N$+$03 also obtained narrow-band CN/TiO photometry in NGC\,147. They
find 146 C stars, of which we detect 102 in DUSTiNGS with $\langle
[3.6]-[4.5] \rangle = -0.01$~mag. The blue colors of some C stars may
be due absorption from CO and/or C$_{3}$ at 4--6~\micron.  Only 7 of
the N$+$03 stars are x-AGB variable stars here: these have a maximum
$[3.6]-[4.5] = 0.72$~mag. Of these 7 variable x-AGB stars, N$+$03
identified 4 as C-rich, 1 as O-rich, and 2 as ``other''.

\citet{BattinelliDemers2004a} also used CN/TiO to identify C stars in
NGC\,147. In addition to those classified by N$+$03, 13 more variable
x-AGB candidates are classified as C-rich.

\citet{Sohn+2006} obtained $JHK$ photometry of stars in NGC\,147 and
classified an additional 10 variable x-AGB stars as C-rich. These
stars have $1.95 < J-K < 3.41$ and $0.16 < [3.6]-[4.5] <
0.79$~mag. Altogether, 27 variable x-AGB stars have been classified as
C-rich based on previous surveys.

\subsubsection{IC\,10}

\citet{Demers+2004} obtained narrow-band CN/TiO photometry of IC\,10
and find 676 C stars, of which 356 are detected here with $\langle
[3.6]-[4.5] \rangle = -0.09$~mag. Only 31 of these 356 C stars fall in
the x-AGB region of the CMD, the reddest of which has
$[3.6]-[4.5]=0.53$~mag. Of the 235 variable x-AGB stars detected here,
13 are classified as C-rich.

IR spectra of 9 dusty O-rich stars in IC\,10 were obtained by
\citet{Lebouteiller+2012} using {\it Spitzer}. None of these sources
are variable in DUSTiNGS. These stars are brighter than the x-AGB
candidates ($m_{3.6} \approx 13$~mag), but they show similar colors to
the x-AGB stars ($0.2 < [3.6]-[4.5] < 0.5$ mag).

\citet{Magrini+2003} identified 16 planetary nebulae in IC\,10.  Only one
(PN7) is potentially detected in the DUSTiNGS images, though blending
with nearby sources caused it to be excluded from the ``good-source''
catalog (see Paper\,I). Because of this, it is not included in the
variability analysis here.

\subsubsection{Other Galaxies}
\label{sec:other}

\begin{description}[leftmargin=1em]

\item[IC\,1613] Carbon stars were identified via CN/TiO narrow-band
photometry by \citet{Albert+2000}. They found 195 C stars, of which 107
are detected here with $\langle [3.6]-[4.5] \rangle = -0.01$~mag and
10 are in the x-AGB region of the CMD. Only one of these C stars is
identified as an x-AGB variable star here.

\item[Pegasus\,dIrr] \citet{BattinelliDemers2000} found 40 C stars
  using narrow-band CN/TiO photometry. 31 are detected here and 3 are
  variable x-AGB candidates.

\item[Aquarius] \citet{BattinelliDemers2000} found 3 C stars with
  CN/TiO photometry. None of these are variable here, but one is in
  the x-AGB region of the CMD. \citet{Gullieuszik+2007} found 10 C
  stars using $JHK$ photometry. We detect 8 of these stars, but none
  are variable in DUSTiNGS. This includes 2 LPV candidates, which have
  $M_{3.6}=-9.67$ and $-8.65$~mag and $[3.6]-[4.5] = 0.62$ and
  $0.3$~mag. Their $J-K$ colors are 2.86 and 2.01~mag, respectively.

\item[Sag\,DIG] \citet{DemersBattinelli2002} identified 16 C stars
  with CN/TiO photometry. We detect 13 here, but only one is a
  variable x-AGB candidate. \citet{Gullieuszik+2007} identified C
  stars via $JHK$ photometry, and 3 are variable x-AGB candidates
  here. These stars have $1.74 < J-K < 4.15$~mag. These 4 stars are
  the first dust-producing high-confidence C stars at ${\rm [Fe/H]} <
  -2$, though spectroscopy is required for definitive classification
  as C rich. There are 11 additional confirmed C stars that are not
  detected as variable here, but do fall within the x-AGB region of
  the CMD (Section~\ref{sec:lowZ}).

\item[WLM] \citet{BattinelliDemers2004c} used CN/TiO photometry to
  identify 149 stars. Of these, we detect 121 in DUSTiNGS and 11
  are variable x-AGB candidates.

\item[Leo\,A] \citet{Magrini+2003} identified a planetary nebula is
  also detected in DUSTiNGS with $\langle [3.6] \rangle =16.72$~mag
  and $\langle [3.6]-[4.5] \rangle =1.14$~mag. This source falls on
  the border of the x-AGB region of the CMD, though is not detected as
  variable.

\end{description}

\section{Discussion}
\label{sec:disc}

\subsection{Changes in the Dust}
\label{sec:changes}

%%%%%%%%%%%%%%%%%%%%%%%%%%%%%%%%%%%%%%%%%%%%%%%%%%%%%%%%%%%%%%%
\begin{figure}
\includegraphics[width=0.95\columnwidth]{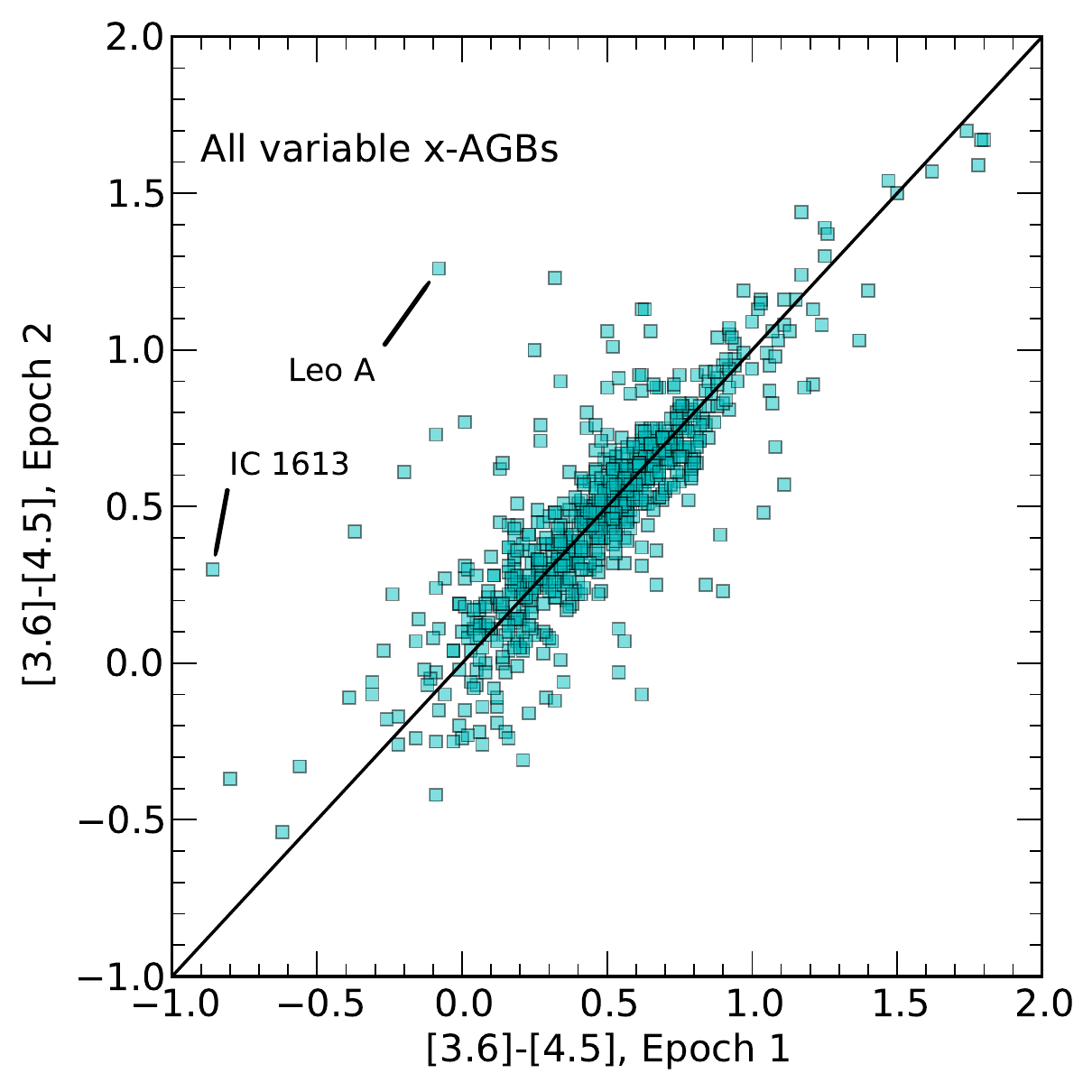}
\caption{Comparison between $[3.6]-[4.5]$ from epoch 1 to epoch 2
  ($\approx$6~month separation) for all variable x-AGB stars. The
  standard deviation in the color change is 0.19~mag. The solid black
  line marks equal colors. \label{fig:colchange}}
\end{figure}
%%%%%%%%%%%%%%%%%%%%%%%%%%%%%%%%%%%%%%%%%%%%%%%%%%%%%%%%%%%%%%%

Changes in the color of a dusty star over time indicate changes in the
temperature and/or the optical depth of the dust. Between the epoch\,1
and epoch\,2 observations, the colors of the x-AGB variable stars do
not change significantly (Fig.~\ref{fig:colchange}), suggesting that
the pulsations within a cycle do not have a strong effect on the dust
on a short timescale.  Comparisons in the color between epoch 0
(Section~\ref{sec:cryo}) and the DUSTiNGS epochs is not possible
because the 3.6 and 4.5~\micron\ observations are not simultaneous in
epoch 0.

A subset of x-AGB variables do show a significant color change between
epochs; two of these stars are marked in
Figure~\ref{fig:colchange}. We cannot disentangle the effects of
temperature and dust mass on the color of these stars with only the
DUSTiNGS data. Nonetheless, it is clear that in some of the x-AGB
variable stars, the circumstellar environment is changing, sometimes
significantly, over a short timescale, implying that dust
formation is not continuous and smooth over the dust-producing phase.

A blue $[3.6]-[4.5]$ color is typical for a less dusty C star that has
CO$+$C$_3$ molecular absorption from 4--6~\micron, but this feature
becomes veiled as the dust emission increases
\citep{vanLoon+08b,Boyer+11}. There are a handful of stars that appear
blue in epoch 1 and red in epoch 2, suggesting that new dust appeared
after the epoch 1 observations. These stars may have undergone an
episode of eruptive dust production, perhaps owing to the influence of
a binary companion. Further monitoring at IR wavelengths could confirm
such phenomena.

%%%%%%%%%%%%%%%%%%%%%%%%%%%%%%%%%%%%%%%%%%%%%%%%%%%%%%%%%%%%%%%
\begin{figure}
\includegraphics[width=\columnwidth]{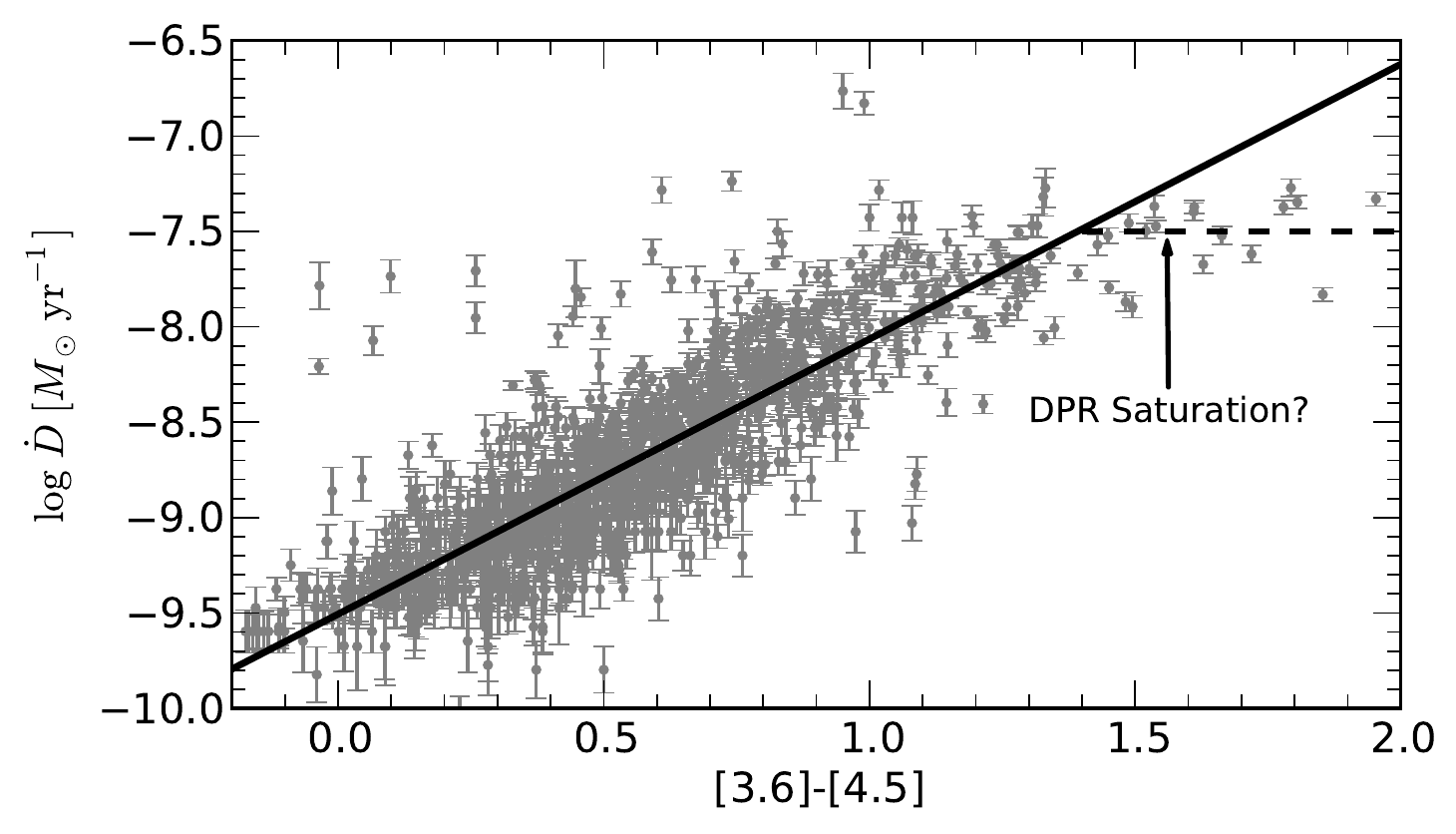}
\caption{Dust-production rates (DPRs) for LMC x-AGB stars, from SED
  fitting to GRAMS models \citep{Riebel+2012}.  The solid line is the
  best fit (Eq.~\ref{eqn:dpr}) and the dashed line marks the possible
  point of saturation in the dust-production rate, which is not
  applied to stars here.\label{fig:grams}}
\end{figure}
%%%%%%%%%%%%%%%%%%%%%%%%%%%%%%%%%%%%%%%%%%%%%%%%%%%%%%%%%%%%%%%

%%%%%%%%%%%%%%%%%%%%%%%%%%%%%%%%%%%%%%%%%%%%%%%%%%%%%%%%%%%%%%%
\begin{figure}
\vbox{
  \includegraphics[width=0.9\columnwidth]{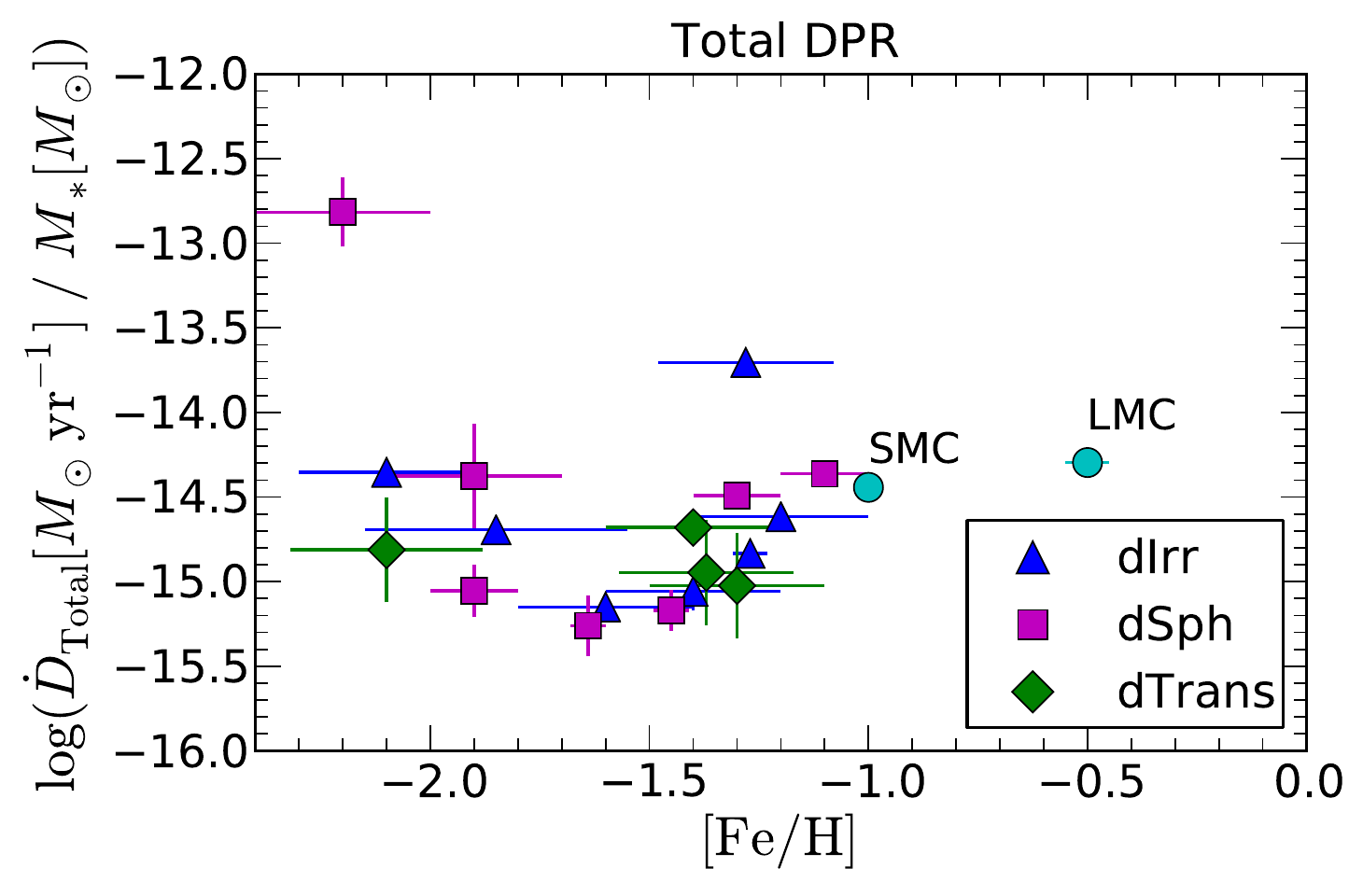}
}
\caption{Total dust-production rate (DPR) normalized to the total
  stellar mass within each galaxy as a function of the galaxy's
  metallicity. Here, we apply the DPR saturation limit shown in
  Figure~\ref{fig:grams} for the reddest stars (also see
  Fig.~\ref{fig:dpr2}). Only variable x-AGB stars are included in the
  DUSTiNGS points. LMC and SMC points were derived using the same
  color-criteria from the SAGE data \citep{Meixner+06,Gordon+11}, and
  include all x-AGB stars. \label{fig:dpr1}}
\end{figure}
%%%%%%%%%%%%%%%%%%%%%%%%%%%%%%%%%%%%%%%%%%%%%%%%%%%%%%%%%%%%%%%

%%%%%%%%%%%%%%%%%%%%%%%%%%%%%%%%%%%%%%%%%%%%%%%%%%%%%%%%%%%%%%%
\begin{figure*}
\hbox{
  \includegraphics[width=0.33\textwidth]{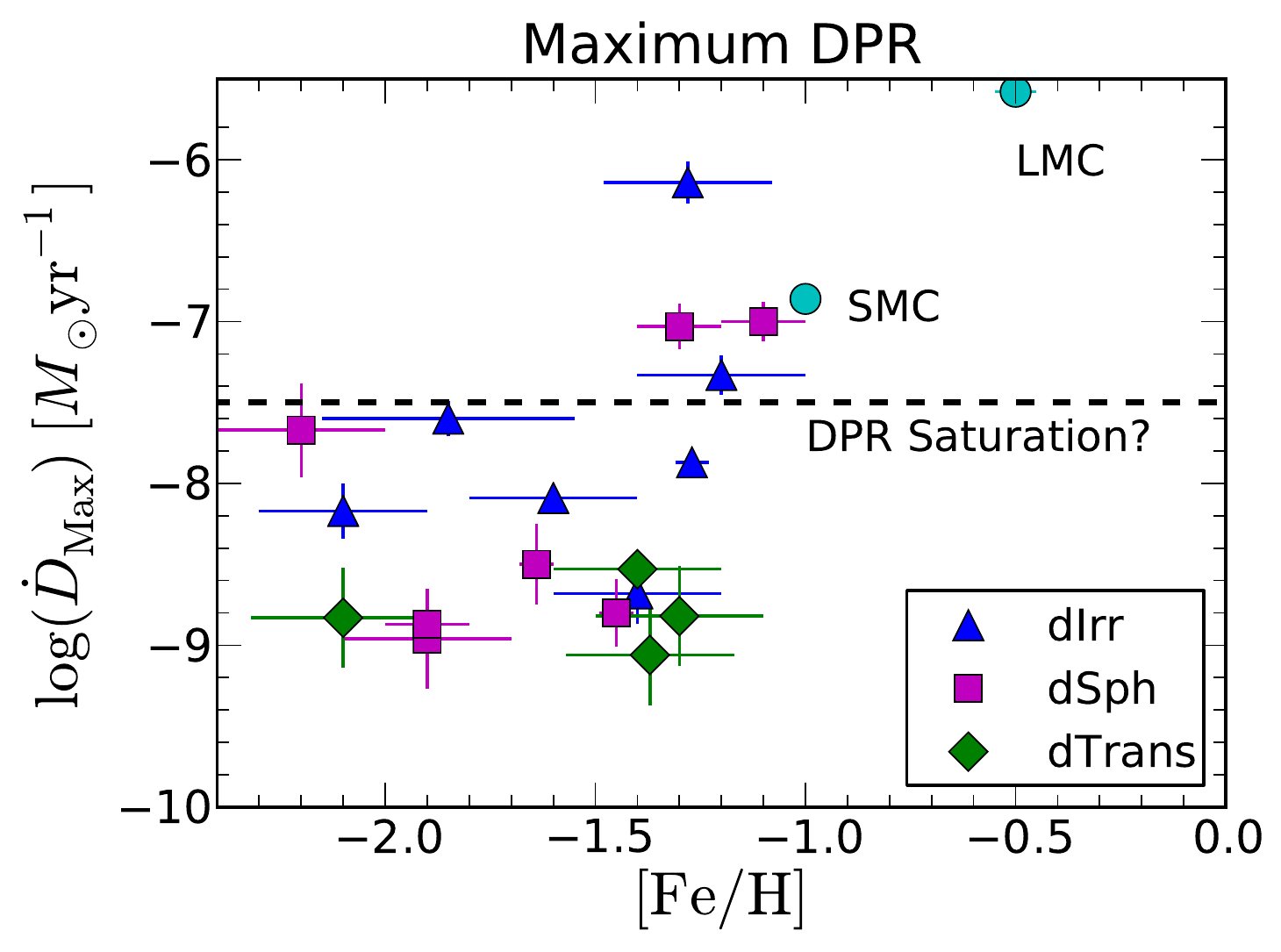}
  \includegraphics[width=0.33\textwidth]{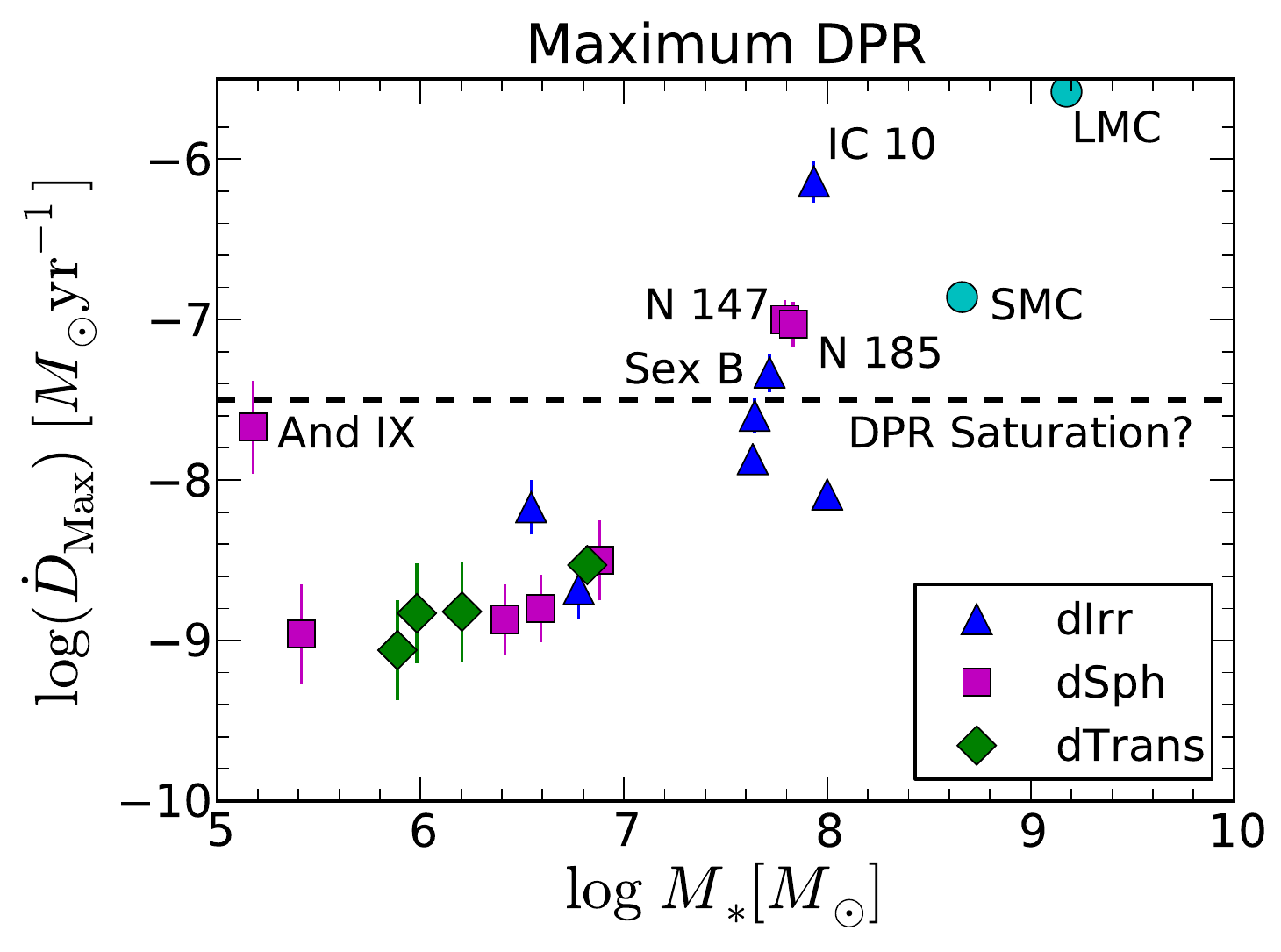}
  \includegraphics[width=0.33\textwidth]{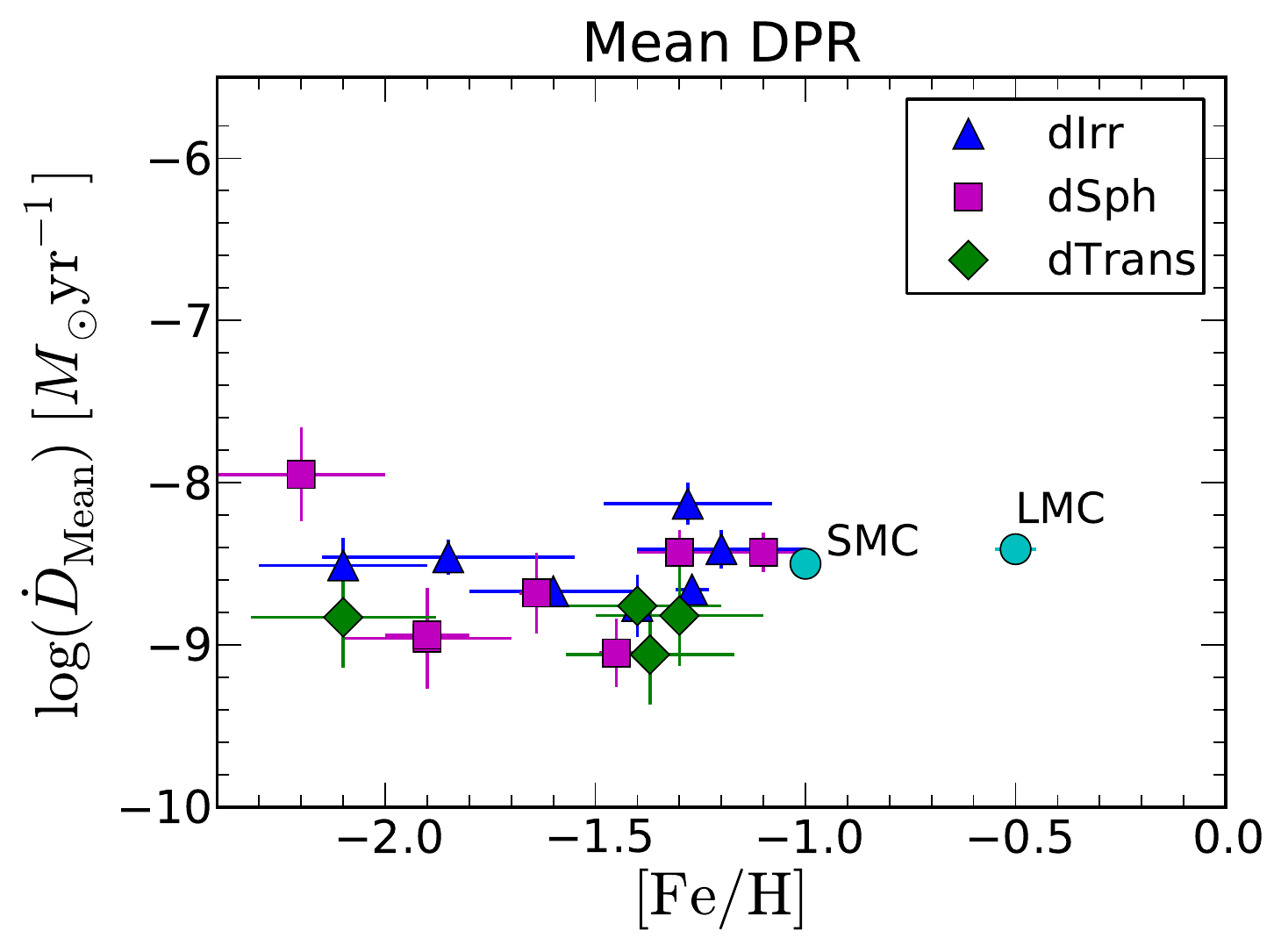}
}
\caption{{\it Left:} Maximum dust-production rate (DPR) among the
  variable x-AGB stars within each galaxy as a function of
  metallicity. The point of possible DPR saturation
  (Fig.~\ref{fig:grams}) is not applied here but is marked by a dashed
  line. {\it Middle:} The 6 galaxies with the largest $\dot{D}_{\rm
    max}$ are also the most massive galaxies in the sample: LMC, SMC,
  IC\,10, NGC\,185, NGC\,147, and Sextans\,B. {\it Right:} Mean DPR
  (with the DPR saturation limit applied to the reddest stars) among
  the variable x-AGB stars within each galaxy, as a function of
  metallicity.\label{fig:dpr2}}
\end{figure*}
%%%%%%%%%%%%%%%%%%%%%%%%%%%%%%%%%%%%%%%%%%%%%%%%%%%%%%%%%%%%%%%

\subsection{Dust Production}
\label{sec:dpr}

\citet{Riebel+2012} measured the dust produced by the entire LMC AGB
population by fitting the full SEDs from the optical to the IR using
the Grid of RSG and AGB modelS
\citep[GRAMS;][]{Sargent+2011,Srinivasan+2011}.  GRAMS assumes a grain
mixture of 90\% amorphous carbon and 10\% SiC with optical constants
from \citet{Zubko+1996} and \citet{Pegourie+1988}, respectively, and a
standard KMH grain size distribution \citep{Kim+1994}. Using the
catalog from \citet{Riebel+2012}, we find that most of stars
with $[3.6]-[4.5]>0.1$~mag are C-rich and that their dust-production
rates ($\dot{D}$) increase with the color as (Fig.~\ref{fig:grams}):

\begin{equation}
\label{eqn:dpr}
\log \dot{D} [M_\odot\,{\rm yr}^{-1}]= -9.5 + [1.4 \times ([3.6]-[4.5])].
\end{equation}

\noindent We apply this relationship to the x-AGB variable star
candidates in the DUSTiNGS galaxies to get a first estimate of their
total dust production, using an average between the epoch 1 and 2
colors (Section~\ref{sec:changes}). For stars redder than $[3.6]-[4.5]
\approx 1.5$~mag, the GRAMS dust-production rate appears to saturate
at a $\log \dot{D} \sim -7.5\ [M_\odot\ {\rm yr}^{-1}]$. We do not
apply this saturation limit to the rates reported in
Table~\ref{tab:varcandy}, but we do apply it in Figure~\ref{fig:dpr1}
and in the right panel of Figure~\ref{fig:dpr2}. If real, this
saturation limit applies to 8 x-AGB variables total in Sextans\,B,
NGC\,147, NGC\,185, and IC\,10 (Fig.~\ref{fig:varcmd}).

We caution that the use of different optical constants and/or the
assumption of a different wind outflow velocity (GRAMS assumes
10~km\,s$^{-1}$) could result in dust-production rates that differ by
up to a factor of 10. Therefore, rates reported here are not
absolutely accurate and should only be compared to rates measured
using similar assumptions. We assume that these parameters do no
change from galaxy to galaxy.

Figure~\ref{fig:dpr1} shows the total dust production, normalized to
the total stellar mass, from the variable x-AGB candidates as a
function of metallicity. For the LMC and SMC points, we used the SAGE
catalog to select x-AGB stars using the same selection criteria used
here for DUSTiNGS galaxies.  We note that the LMC and SMC points
include the entire x-AGB population, while the DUSTiNGS points include
only the x-AGB stars detected as variable (Section~\ref{sec:prob}).
The total dust output shows significant scatter and does not appear to
be strongly affected by metallicity.  The magenta square in the upper
left corner belongs to And\,IX, and is totally dominated by a single
x-AGB variable. This star (\#21181) is marginally affected by imaging
artifacts (Section~\ref{sec:xagb}), so its dust-production rate may be
inflated.  Except for this And\,IX star, no other x-AGB stars that may
be affected by imaging artifacts are included in this plot or in
Figure~\ref{fig:dpr2}.

In Figure~\ref{fig:dpr2}, each point indicates the mean (or the
maximum) dust-production rate among the variable x-AGB stars within
one of the galaxies.  While the mean $\dot{D}$ is similar at all
metallicities, the maximum $\dot{D}$ may show a slight preference for
more metal-rich galaxies. However, the five galaxies with the highest
maximum $\dot{D}$ (LMC, SMC, IC\,10, NGC\,147, and NGC\,185) all have
the largest x-AGB populations, so are less affected by stochastics and
are therefore more likely to harbor the rarest, very dusty x-AGB
stars.

\subsection{Dust at Very Low Metallicity}
\label{sec:lowZ}

%%%%%%%%%%%%%%%%%%%%%%%%%%%%%%%%%%%%%%%%%%%%%%%%%%%%%%%%%%%%%%%
\begin{figure}
\vbox{
\includegraphics[width=\columnwidth]{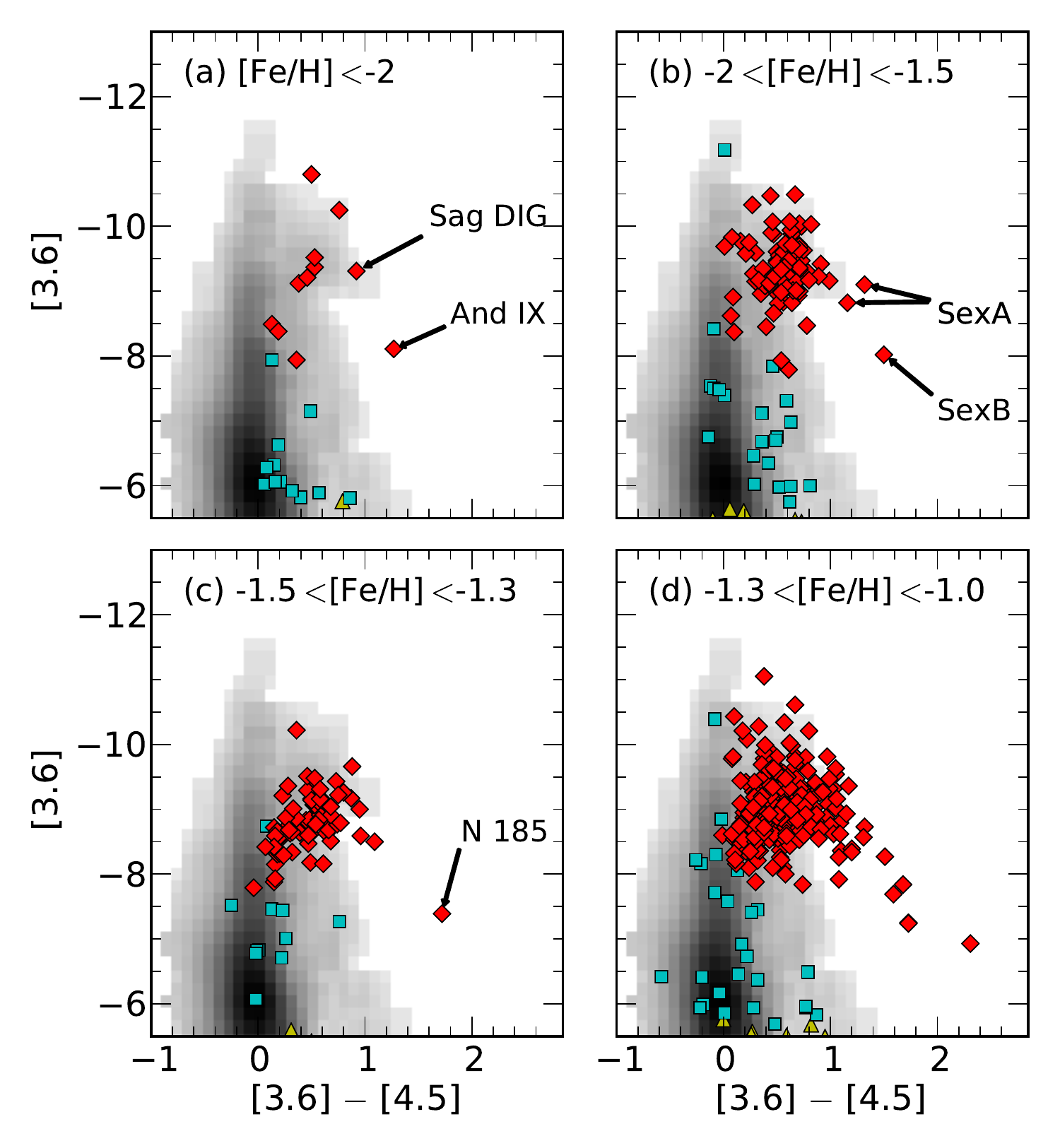}
}
\caption{Location of variable sources on the CMD, separated by ${\rm
    [Fe/H]}$.  The x-AGB variable star candidates are plotted as red
  diamonds, AGB star candidates as cyan squares, and other variables
  as yellow triangles. The greyscale background in all panels is the
  mean background-subtracted CMD (plotted as the density of sources)
  for all 50 target galaxies. All x-AGB
  candidates in {\it (d)} redder than $[3.6]-[4.5] = 1.5$~mag are from
  NGC\,147 and IC\,10.\label{fig:varcmd_byZ}}
\end{figure}
%%%%%%%%%%%%%%%%%%%%%%%%%%%%%%%%%%%%%%%%%%%%%%%%%%%%%%%%%%%%%%%

%%%%%%%%%%%%%%%%%%%%%%%%%%%%%%%%%%%%%%%%%%%%%%%%%%%%%%%%%%%%%%%
\begin{figure}
\includegraphics[width=0.85\columnwidth]{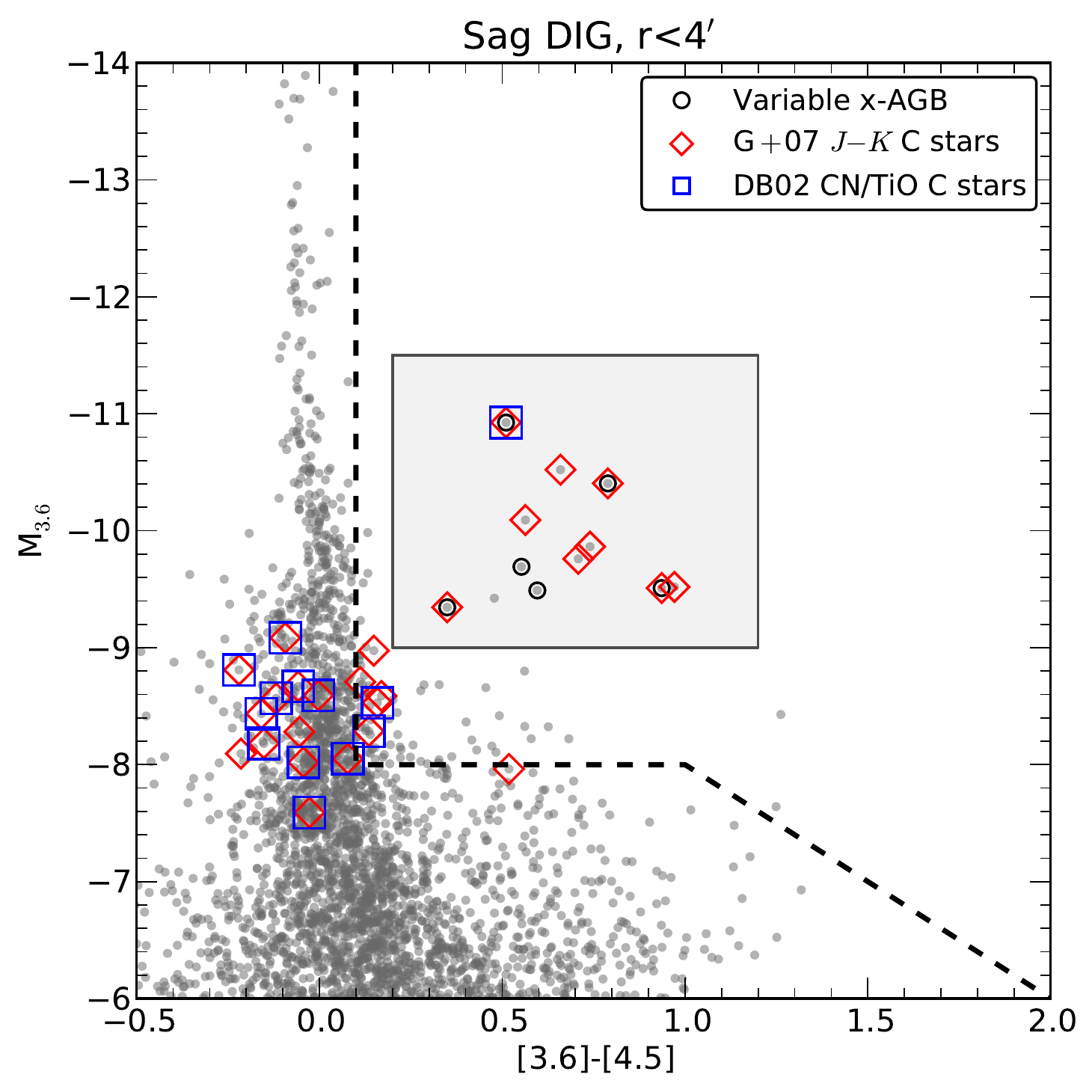}
\caption{CMD of the central 4\arcmin\ of Sag\,DIG showing high
  confidence x-AGB candidates within the grey box. Those detected as
  variable are circled in black. Sources that are confirmed C stars
  based on CN/TiO narrow-band photometry are marked with blue boxes
  \citep[DB02;][]{DemersBattinelli2002}.  Sources that are likely C
  stars based on their $J-K$ colors are marked with red diamonds
  \citep[G$+$07;][]{Gullieuszik+2007}. All sources to the right of the
  black dashed line are x-AGB candidates.\label{fig:sagdig}}
\end{figure}
%%%%%%%%%%%%%%%%%%%%%%%%%%%%%%%%%%%%%%%%%%%%%%%%%%%%%%%%%%%%%%%

\begin{deluxetable}{lrrrrr}
\tablewidth{\columnwidth} 
\tabletypesize{\footnotesize}
\tablecolumns{6} 
\tablecaption{Additional x-AGB candidates with ${\rm [Fe/H]}<-2$\label{tab:sagdig}}

\tablehead{
\colhead{}&
\colhead{GSC ID}&
\colhead{RA}&
\colhead{Dec}&
\colhead{$\langle [3.6] \rangle$}&
\colhead{$\langle [4.5] \rangle$}\\
\colhead{Galaxy}&
\colhead{}&
\colhead{(J2000)}&
\colhead{(J2000)}&
\colhead{(mag)}&
\colhead{(mag)}
}

\startdata
Sag\,DIG   &  34625  &19 30 03.25  &$-$17 41 33.6 &  15.84 &  14.86\\
Sag\,DIG   &  36867  &19 30 02.01  &$-$17 40 19.3 &  15.49 &  14.75\\
Sag\,DIG   &  38522  &19 30 01.11  &$-$17 39 59.1 &  15.59 &  14.88\\
Sag\,DIG   &  41056  &19 29 59.76  &$-$17 41 04.9 &  15.26 &  14.70\\
Sag\,DIG   &  44334  &19 29 57.94  &$-$17 40 17.3 &  14.82 &  14.16\\
Sag\,DIG   &  50071  &19 29 54.74  &$-$17 41 04.4 &  15.94 &  15.44\\
  LGS\,3   &  68695  &01 03 43.01  &$+$21 50 45.4 &  14.73 &  14.11

\enddata

\tablecomments{\ High likelihood x-AGB candidates in Sag\,DIG and
  LGS\,3 that are not detected as variable in DUSTiNGS, but are located in the
  shaded region of Fig.~\ref{fig:sagdig}.}
\end{deluxetable}

In globular clusters, dust-producing AGB stars have potentially been
detected down to ${\rm [Fe/H]} =-2.4$
\citep[e.g.,][]{Boyer+06,Boyer+08,Boyer+09a,McDonald+09,McDonald+2011b,McDonald+2011a,Sloan+2010}. These
stars are low-mass ($M \lesssim 1~M_\odot$), oxygen-rich stars that
will not ultimately contribute much to the total dust budget of a
galaxy \citep[e.g.,][]{Zhukovska+2013,Schneider+2014}. Searches for
more massive metal-poor dust-producing AGB stars have yielded few
examples in Local Group dwarf galaxies \citep{Sloan+2012}. The most
metal-poor example is Mag\,29 in Sculptor dSph \citep[${\rm [Fe/H]} =
  -1.68$ and $\log \dot{D}\ {[M_\odot\,{\rm yr}^{-1}]} =
  -8.21$;][]{Sloan+09}, though \citet{Sloan+2012} estimate that its
metallicity may be as high as ${\rm [Fe/H]} = -1$. We
  find a total of 111 variable x-AGB stars with ${\rm [Fe/H]}<-1.5$
  and 12 with ${\rm [Fe/H]} < -2$, suggesting that carbon formed in
  and around the stellar core is successfully dredged up and condensed
  into carbon grains at all metallicities.

We assume here that the wind expansion velocity and the dust
properties are the same in all the DUSTiNGS galaxies, but this
assumption may not be valid. For example, \citet{Wachter+2008} find
that the expansion velocity decreases from solar metallicity to SMC
metallicity when the ratio C/O is fixed in their models. However,
\citet{Wachter+2008} also find that when models are given the same
absolute abundance of free carbon, the SMC-, LMC-, and
solar-metallicity models have similar expansion velocities and
dust-condensation efficiency. The DUSTiNGS findings support this
scenario.

We show the CMD of x-AGB variable star candidates as a function of
metallicity in Figure~\ref{fig:varcmd_byZ}. The sources in Sag\,DIG,
And\,IX, and LGS\,3 (${\rm [Fe/H]}<-2$) are of particular interest, as
they indicate that AGB stars can produce large amounts of dust even in
the early, metal-poor Universe. The AGB models from \citet{Marigo+13}
indicate that C stars can form at least as early as $10^{8}$~yr at
these metallicities ($M_{\rm initial} \sim 5~M_\odot$), and perhaps
even earlier depending on the details of the interplay between
dredge-up and hot-bottom burning. AGB stars must therefore contribute
significantly to the large dust reservoirs discovered in distant
quasars \citep{Bertoldi+2003,Robson+2004,Beelen+2006}. 

Sag\,DIG contains the largest very metal-poor x-AGB population (${\rm
  [Fe/H]} = -2.1$), with 17 stars that are either x-AGB variables or
confirmed C stars within the x-AGB region of the CMD
(Section~\ref{sec:other}). There is one additional unconfirmed source
that nonetheless has colors similar to the confirmed x-AGB stars and
is within 4\arcmin\ of the galaxy center (Fig.~\ref{fig:sagdig}). We
consider this source to be a likely x-AGB star. Similarly,
LGS\,3 contains one additional possible x-AGB star along with
the one that was identified as variable. These likely x-AGB stars are
listed in Table~\ref{tab:sagdig}.

No variable x-AGB stars were discovered in the most metal-poor
DUSTiNGS galaxies (${\rm [Fe/H]}<-2.2$: UMa\,II, Segue\,I, Leo\,IV,
Coma\,Berenices, Bootes\,I, or Hercules). These galaxies have very low
masses ($M_{\rm V}>-6.6$~mag; for comparison And\,IX has $M_{\rm
  V}=-8.1$~mag) and show no evidence of recent star formation
\citep[Paper\,I and][]{McConnachie+2012}, so it is not surprising that
they are lacking examples of intermediate-aged AGB stars in
general. In Paper\,I, we estimated the size of the AGB population
after subtraction of foreground and background sources.  In Hercules,
we estimated a total of $20\pm9$ AGB stars, which is the smallest
detected AGB population among the DUSTiNGS galaxies. In the other 4
very metal-poor galaxies, we found only upper limits.

\subsection{Implications for Dust in the Interstellar Medium}

If AGB stars can efficiently produce dust at any metallicity as the
DUSTiNGS data suggests, it follows that the dust-to-gas ratios (DGRs)
in the (C-rich) stellar envelopes may be similar at all metallicities
when the stars are in the dust-producing ``superwind'' phase
\citep[][]{Habing+96,Groenewegen+07}. This implies that the DGR in the
interstellar medium (ISM) should also be independent of metallicity if
AGB stars are the dominant source of interstellar dust and dust grains
are minimally processed after leaving the circumstellar envelope.
Observations of nearby galaxies show a clear correlation between
metallicity and DGR in the ISM
\citep[e.g.,][]{Galametz+2011,Sandstrom+2013,Fisher+2014}, suggesting
that AGB dust may be quickly destroyed or shattered
\citep[cf.][]{Temim+2014}. The remnants of grain
destruction/shattering may then provide the seeds for grain accretion
in molecular clouds.

\section{Conclusions}
\label{sec:concl}

We use 2-epoch 3.6 and 4.5~\micron\ photometry to identify 710
variable sources in a sample of 50 nearby dwarf galaxies that were
observed by the DUSTiNGS survey (Paper\,I). Among these variable
sources, 526 have $[3.6]-[4.5]>0.1$~mag colors that roughly correspond
to the ``extreme'' (x-)AGB stars detected in the Magellanic Clouds,
which have been shown to dominate the total AGB dust production in
those galaxies.  We find these x-AGB variable stars in 32 of the
DUSTiNGS galaxies, with metallicities ranging from $-2.2 < {\rm
  [Fe/H]} < -1.1$. Previous optical and near-IR surveys classified 70 of these
x-AGB variable stars as confirmed C stars.

Of the 526 x-AGB variables identified, 12 are in galaxies with ${\rm
  [Fe/H]} < -2.0$. These include 4 known C stars in the very
metal-poor galaxy Sag\,DIG (${\rm [Fe/H]} = -2.1$), which we have
found to be the most metal-poor confirmed dust-producing C stars
known.

The 2-epoch minimum amplitudes of the variable x-AGB candidates
indicate an increase in amplitude with color, supporting a link between
dust production and pulsation. The mean amplitudes show no variation
with metallicity.

We use the $[3.6]-[4.5]$ color to estimate the dust-production rates
of the x-AGB candidates, and we find that the mean dust-production
rate among the x-AGB stars within any galaxy is independent of
metallicity. This trend holds even when including the Large and Small
Magellanic Clouds, suggesting that dust production occurs in C stars
with similar efficiency at any metallicity down to at least ${\rm
  [Fe/H]} = -2.2$. The maximum dust-production rate within a given
galaxy may show a trend with metallicity. However, this trend is more
likely due to a sample bias wherein the most massive galaxies that are
more likely to contain the rarest dusty sources are also the galaxies
with the highest metallicities.

\acknowledgements

We thank Patricia Whitelock \& Michael Feast for discussions about
stellar variability that improved the paper and the referee for
helpful comments. This work is supported by {\it Spitzer} via grant
GO80063 and by the NASA Astrophysics Data Analysis Program grant
number N3-ADAP13-0058. MLB is supported by the NASA Postdoctoral
Program at the Goddard Space Flight Center, administered by ORAU
through a contract with NASA. RDG was supported by NASA and the United
States Air Force. AZB acknowledges funding by the European Union
(European Social Fund) and National Resources under the ``ARISTEIA''
action of the Operational Programme ``Education and Lifelong
Learning'' in Greece.

%%%%%%%%%%%%%%%%%%%%%%%%%%%%%%%%%%%%%%%%%%%%%%%%%%%%%%%%%%%%%%%
%%%% Bibliography
%%%%%%%%%%%%%%%%%%%%%%%%%%%%%%%%%%%%%%%%%%%%%%%%%%%%%%%%%%%%%%%%
%\bibliographystyle{astroads}%../jphysicsB}
%\bibliography{myrefs,galinfo,inprep}

\appendix
\section{Detection Probabilities of Other Variable Sources}

\begin{figure}
\vbox{
\includegraphics[width=0.5\textwidth]{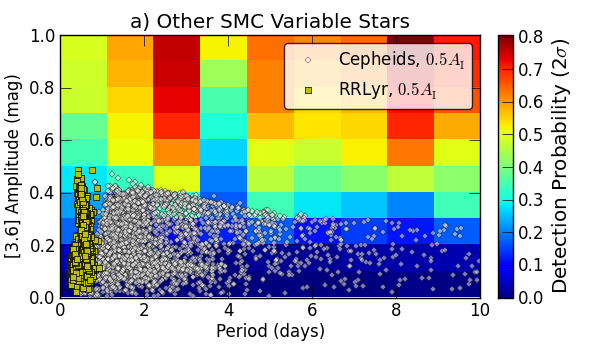}
\includegraphics[width=0.5\textwidth]{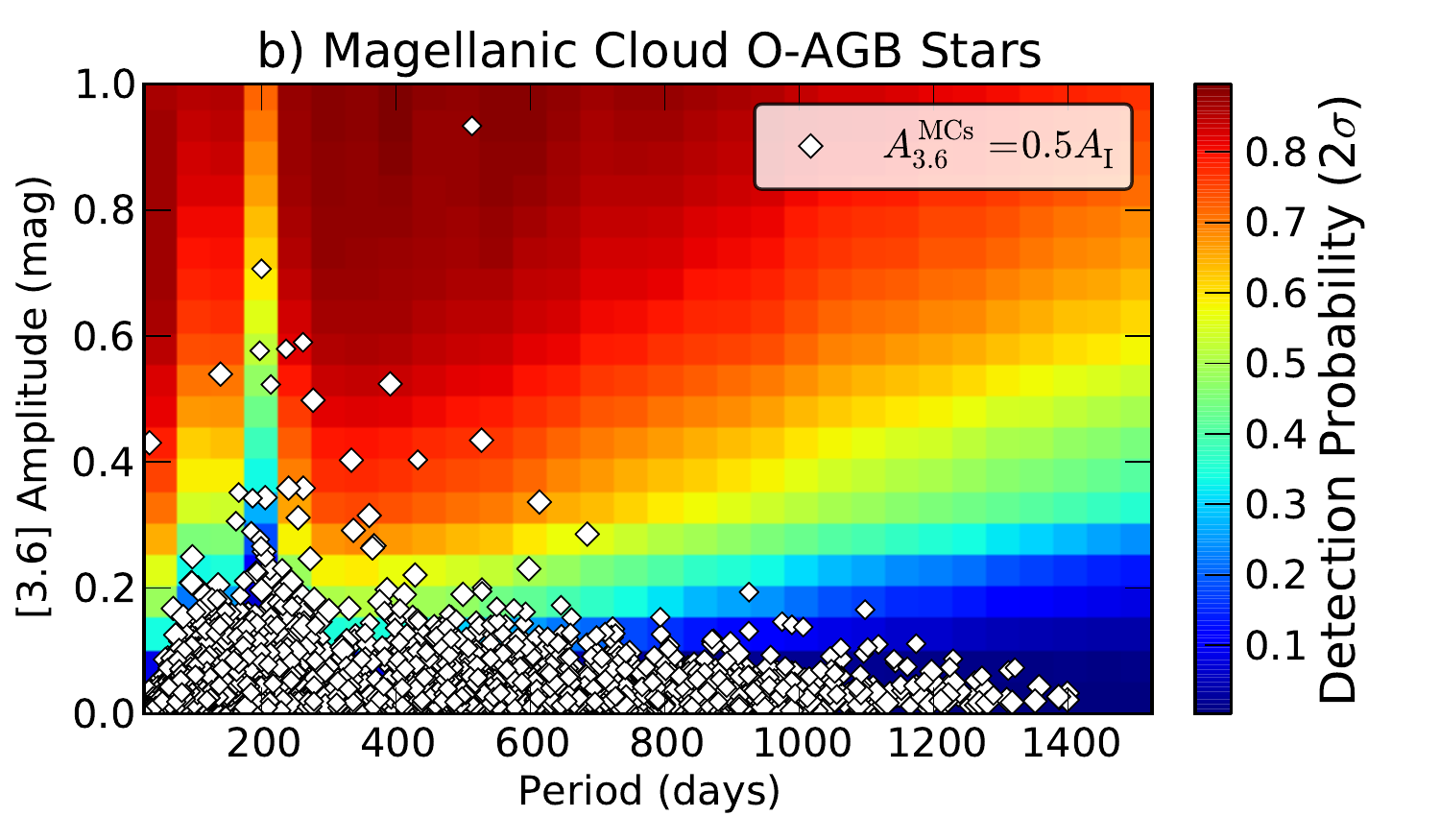}
\includegraphics[width=0.5\textwidth]{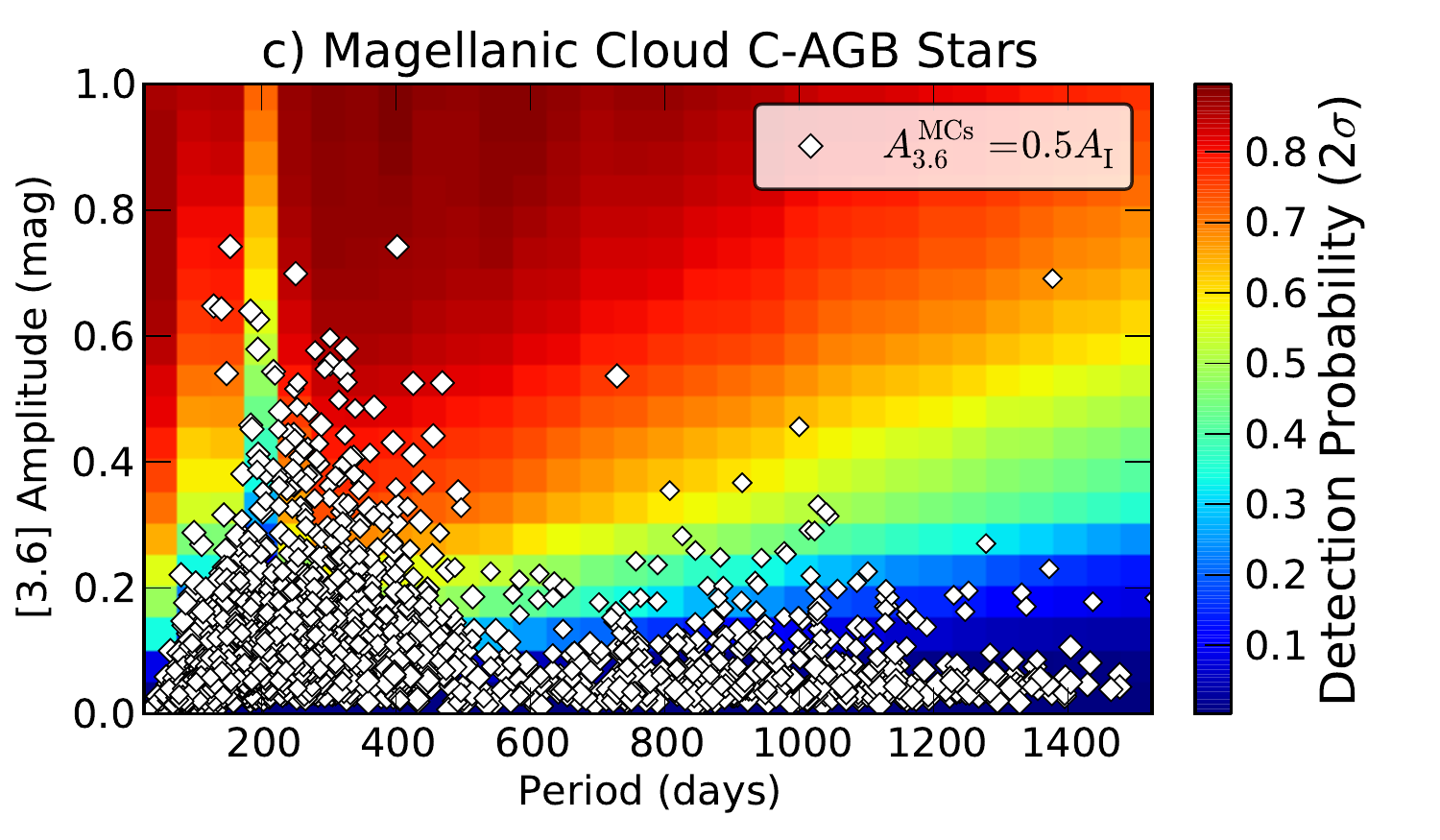}
}
\caption{Probability of detecting {\it a)} Cepheid and RR\,Lyr
  variable stars, {\it b)} O-rich AGB stars with $[3.6]-[4.5] <
  0.1$~mag, and {\it c)} C-rich AGB stars with $[3.6]-[4.5]<0.1$~mag.
  In all three panels, the white and yellow points are variables
  detected by the OGLE survey \citep{Soszynski+08} in the Magellanic Clouds by
  the $I$-band amplitudes and periods. The distribution in the
  detection probability in panels {\it b)} and {\it c)} is identical
  to that in Figure~\ref{fig:prob}. AGB identification is from the
  SAGE Survey \citep{Boyer+11}. Cepheids and RR\,Lyr
  shown in {\it a)} are included regardless of their brightness; most
  stars of this type are fainter than the TRGB.
\label{Afig:sim}}
\end{figure}

Figure~\ref{Afig:sim} shows the detection probability
(Section~\ref{sec:prob}) of other types of variable stars with
DUSTiNGS.

%%%%%%%%%%%%%%%%%%%%%%%%%%%%%%%%%%%%%%%%%%%%%%%%%%%%%%%%%%%%%%%%%%%%%
%%%%%%%%%%%%%%%%%%%%%%%%%%%%%%%%%%%%%%%%%%%%%%%%%%%%%%%%%%%%%%%%%%%%%

\section{Spatial Distribution of variable x-AGB candidates}

\begin{figure}
\vbox{
\includegraphics[width=0.33\textwidth]{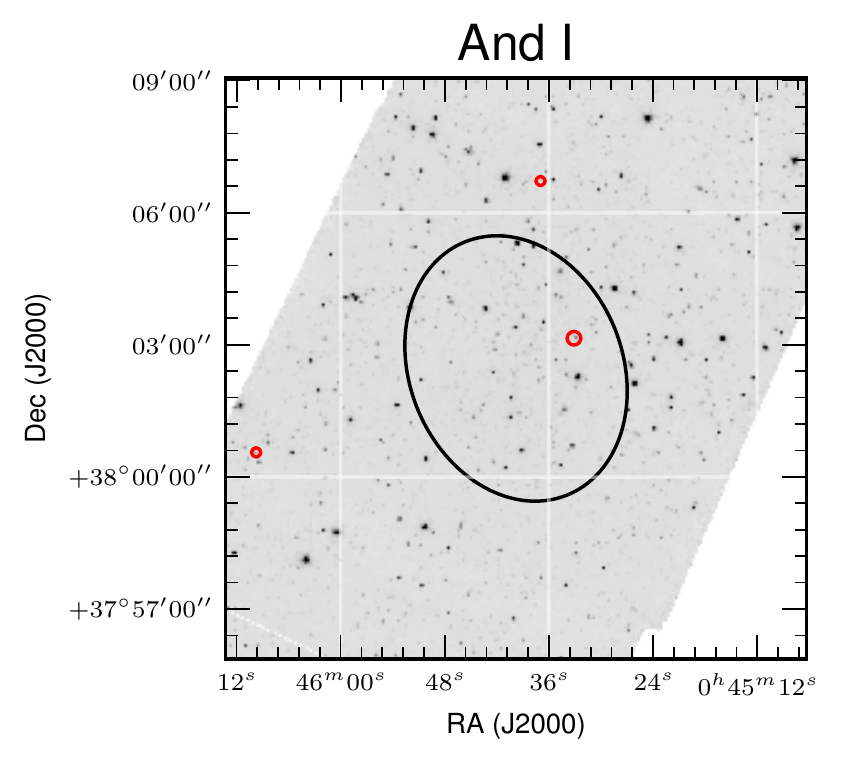}
\includegraphics[width=0.33\textwidth]{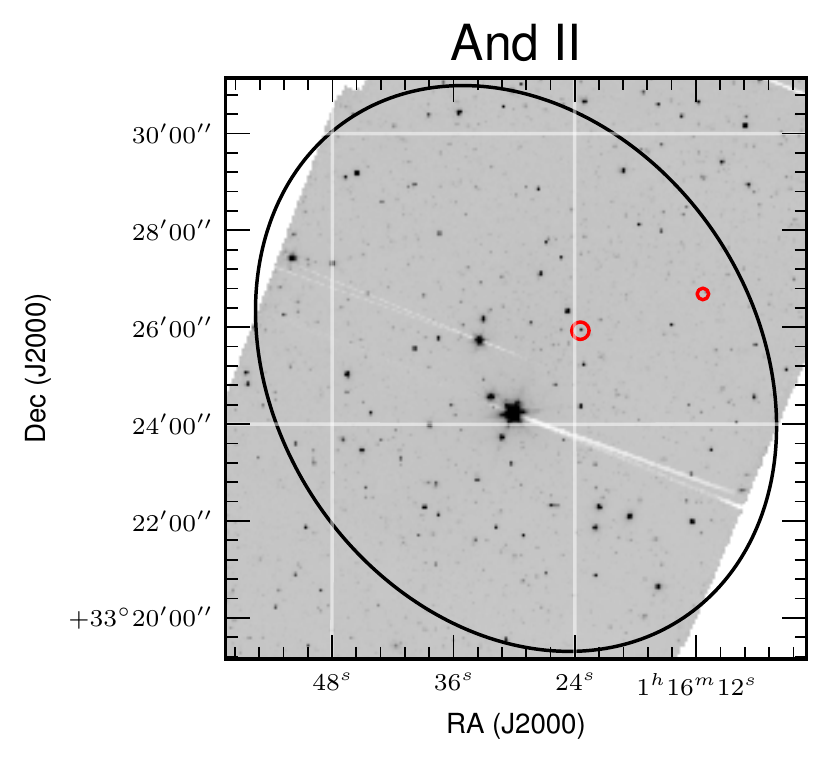}
\includegraphics[width=0.33\textwidth]{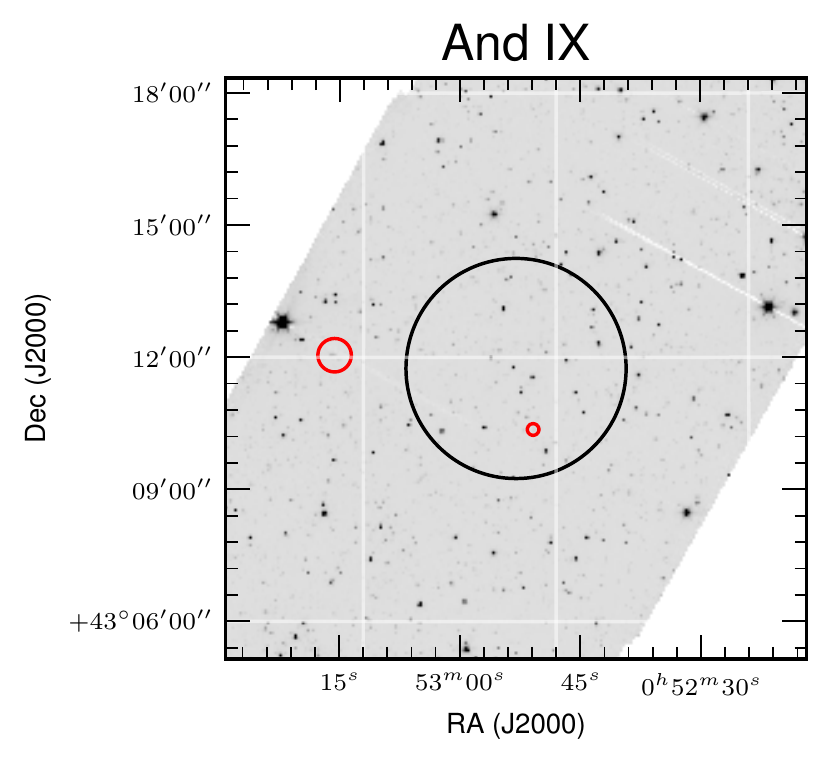}
\includegraphics[width=0.33\textwidth]{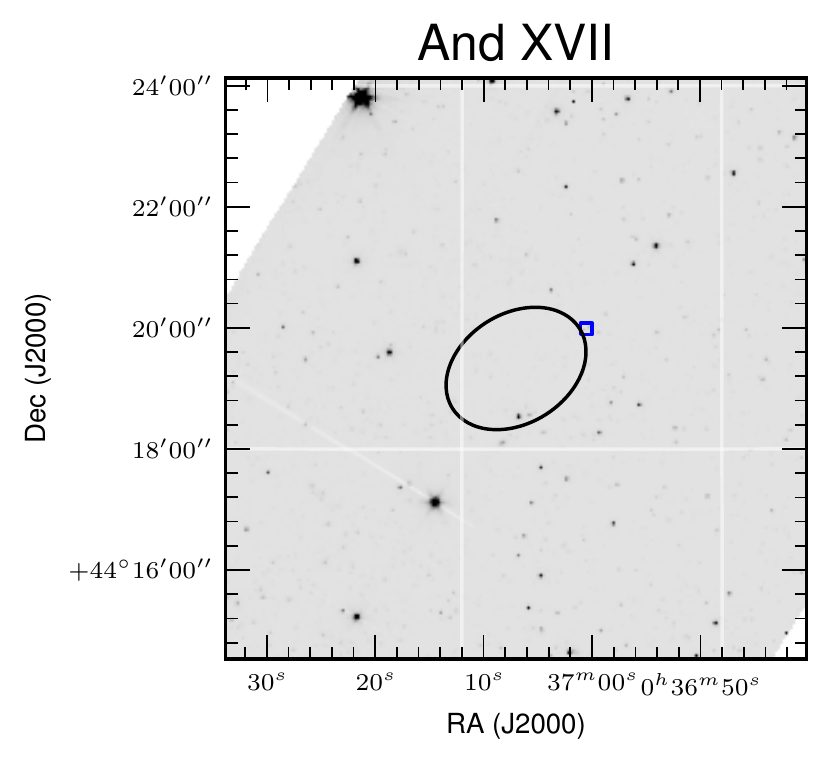}
\includegraphics[width=0.33\textwidth]{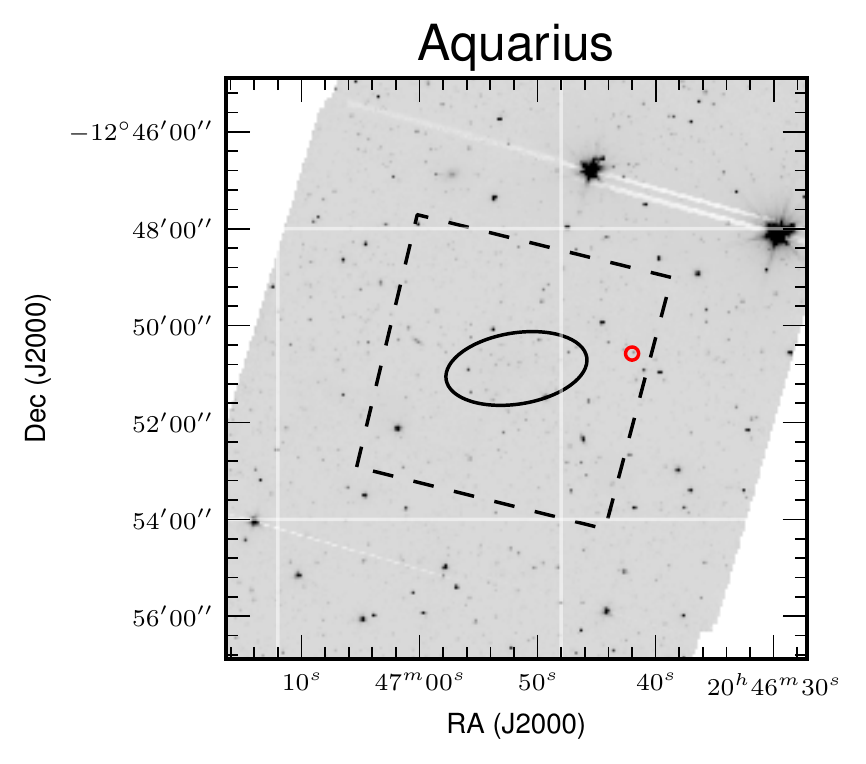}
\includegraphics[width=0.33\textwidth]{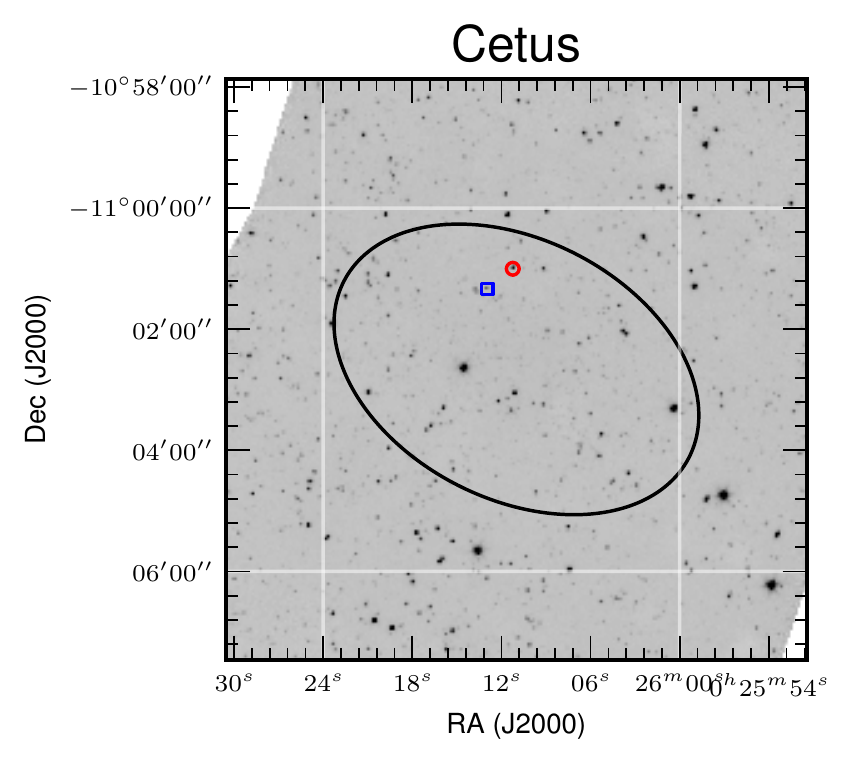}
\includegraphics[width=0.33\textwidth]{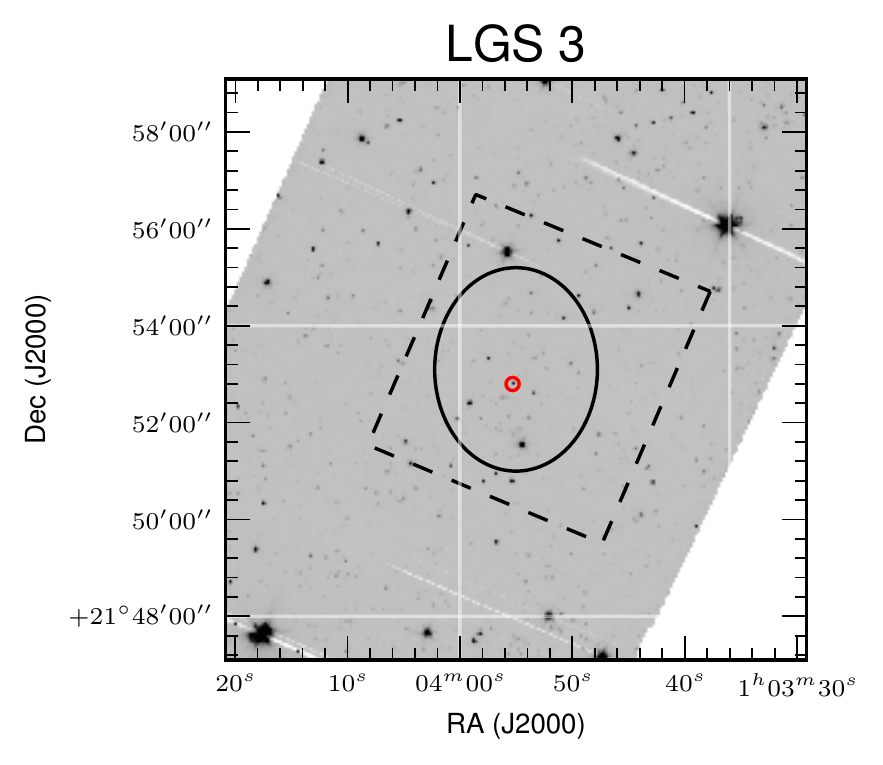}
\includegraphics[width=0.33\textwidth]{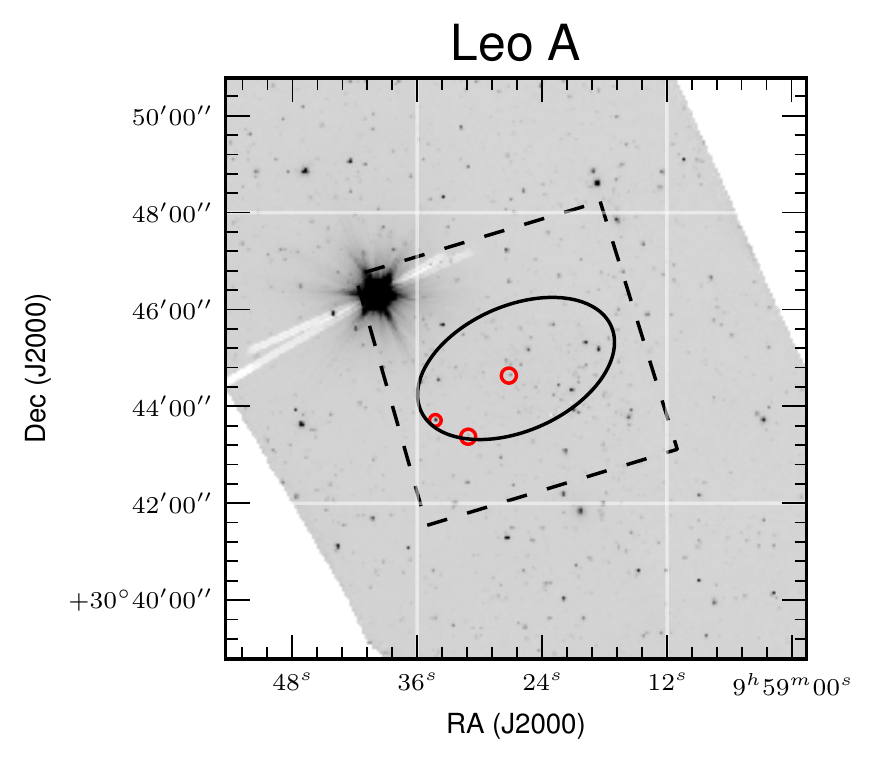}
\includegraphics[width=0.33\textwidth]{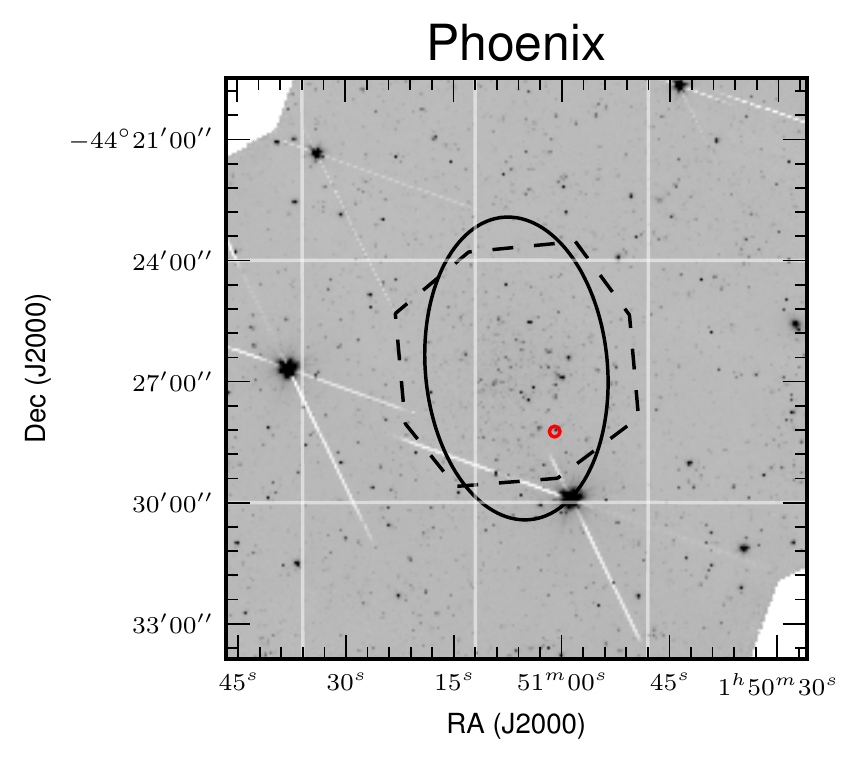}
\includegraphics[width=0.33\textwidth]{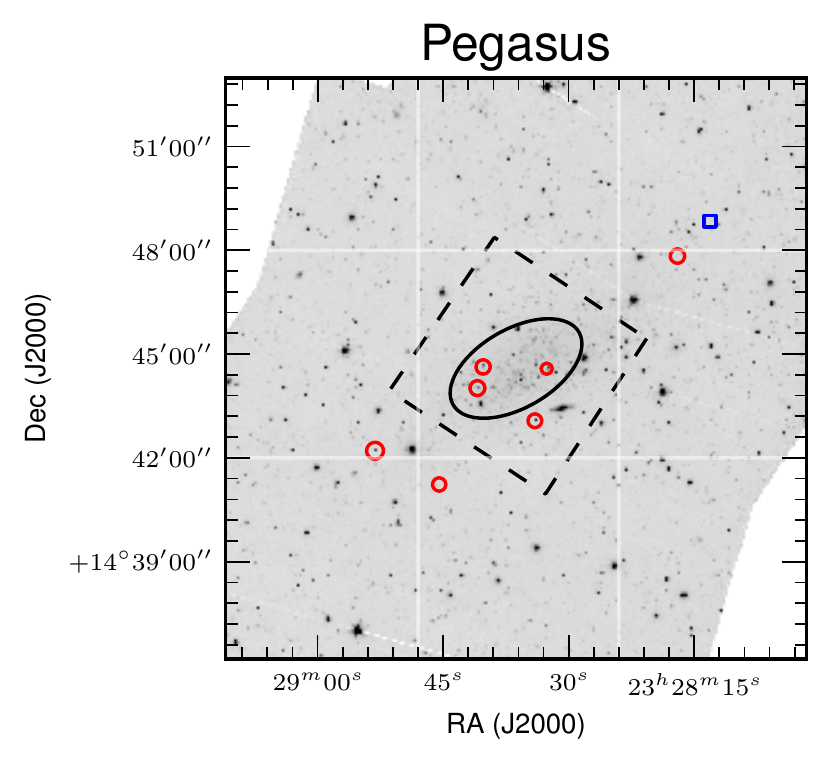}
\includegraphics[width=0.33\textwidth]{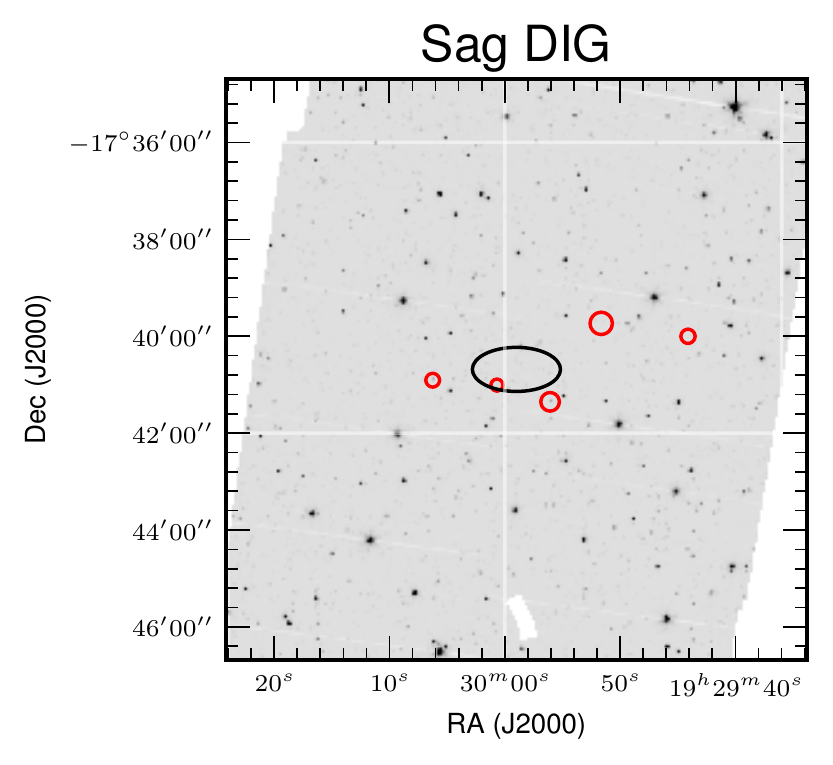}
}
\caption{Spatial location of x-AGB variable candidates. Red circles
  are 3-$\sigma$ variables, blue squares are 2-$\sigma$ variables. The
  size of the symbol is representative of the $[3.6]-[4.5]$ color,
  with larger symbols representing redder colors. The half-light radii
  are marked with black ellipses. And\,IX has unknown ellipticity
  \citep{McConnachie+2012}, so we plot a circular half-light
  radius. Dashed lines marks the epoch 0 coverage. The x-AGB stars that
  may be affected by imaging artifacts (Table~\ref{tab:varcandy}) are not
  included except the very red star (\#21181) in And\,IX.}
\label{fig:spat1}
\end{figure}

\begin{figure}
\vbox{

\includegraphics[width=0.45\textwidth]{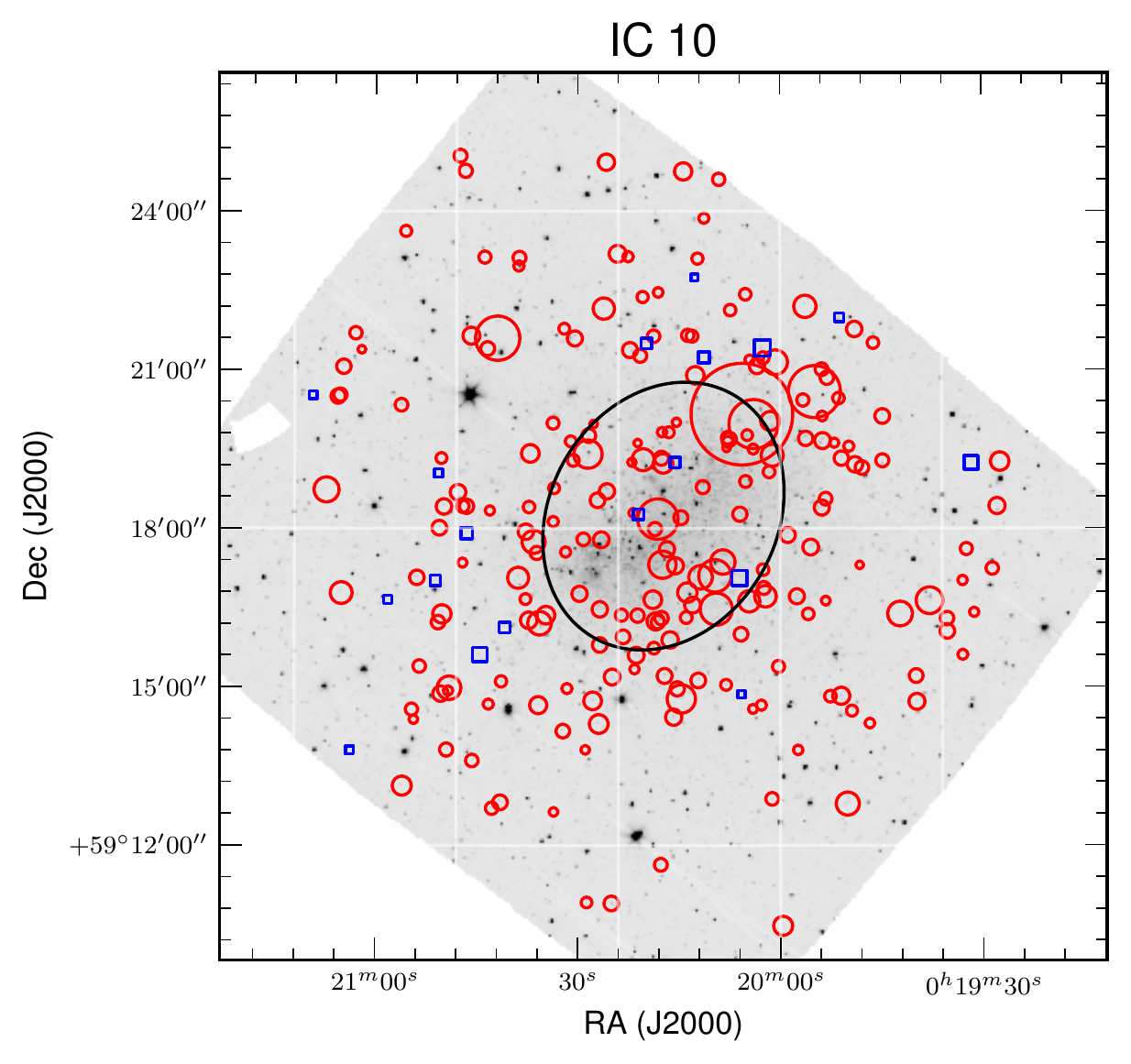}
\includegraphics[width=0.45\textwidth]{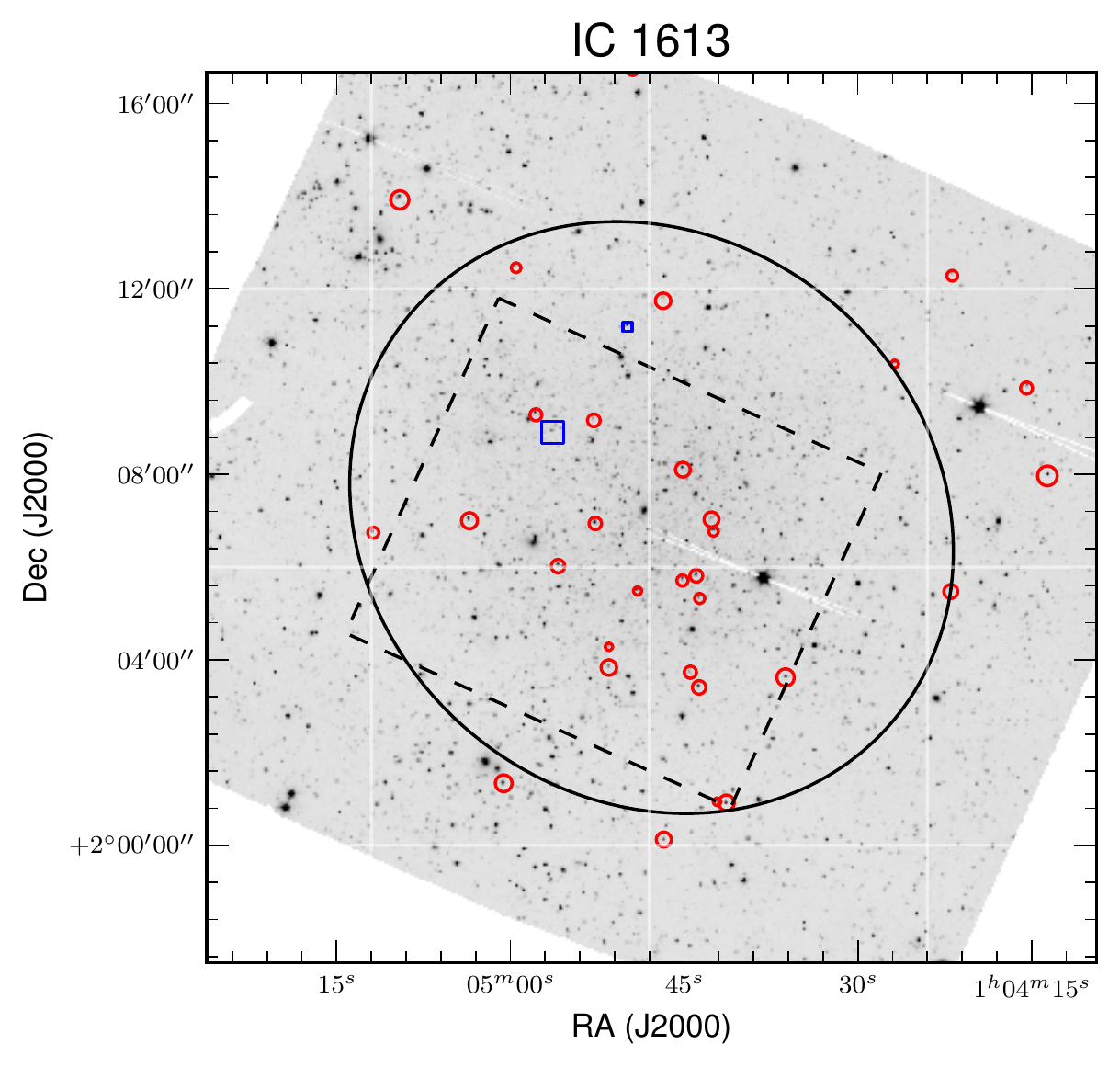}
\includegraphics[width=0.45\textwidth]{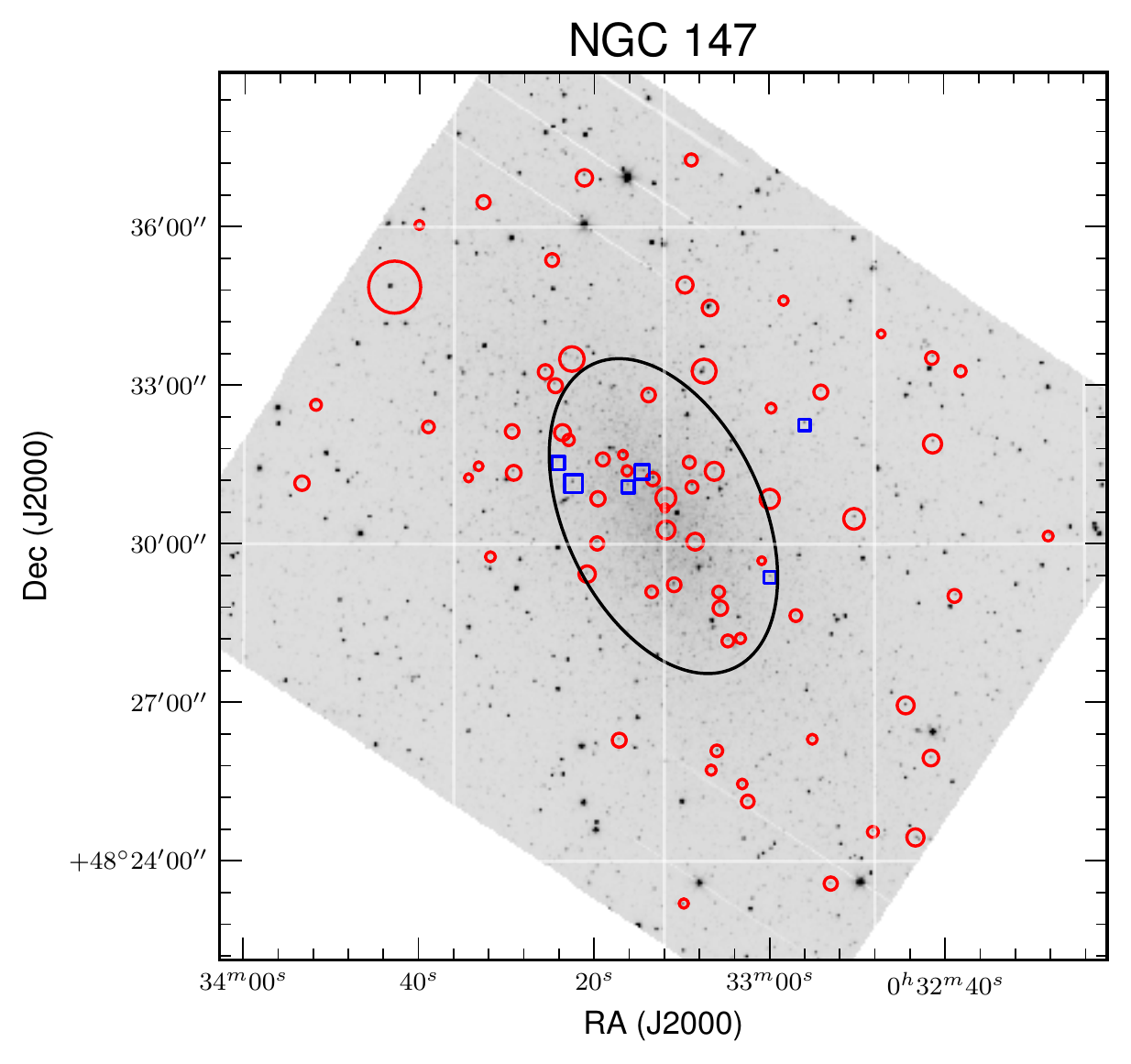}
\includegraphics[width=0.45\textwidth]{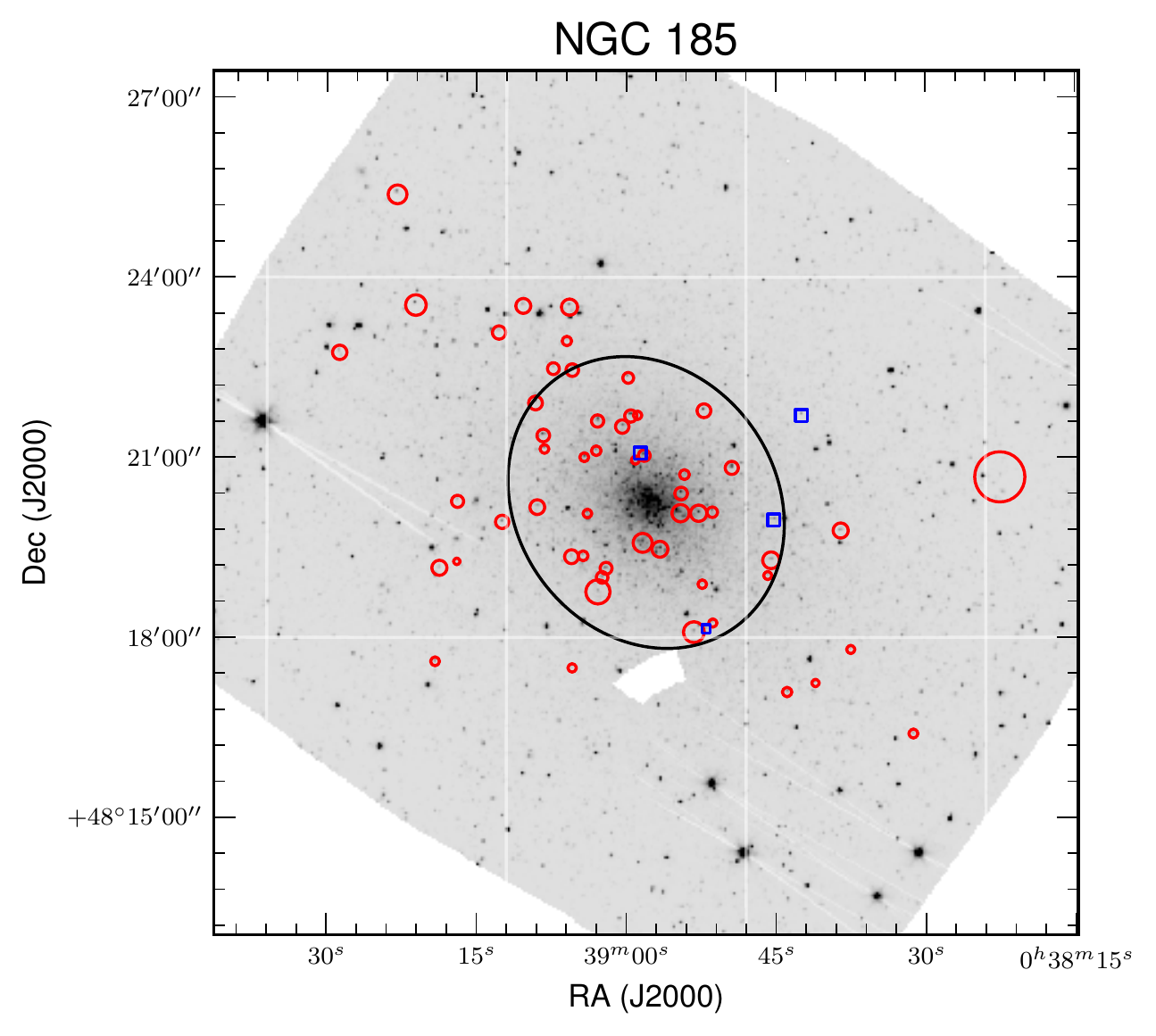}
\includegraphics[width=0.45\textwidth]{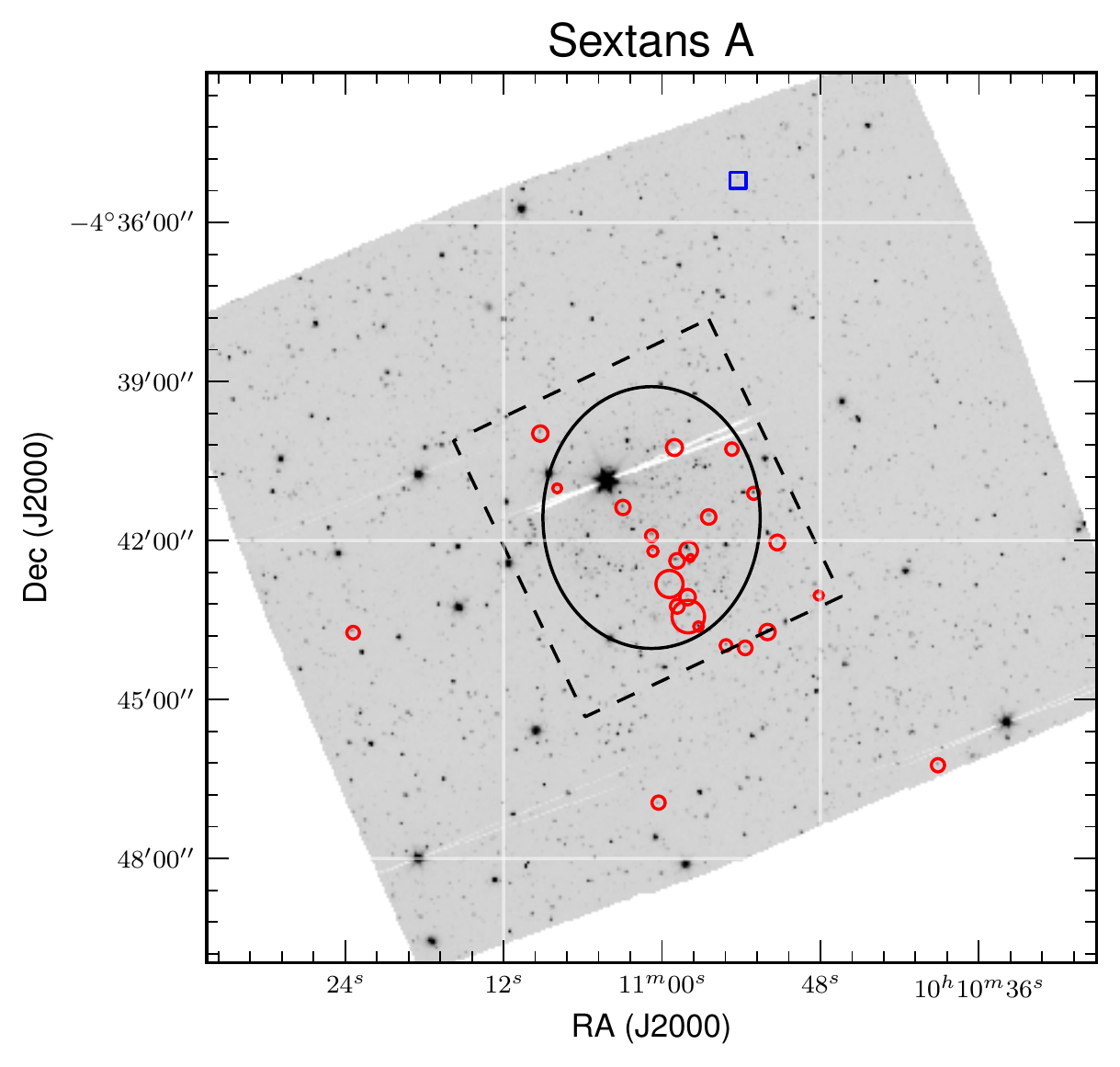}
\includegraphics[width=0.45\textwidth]{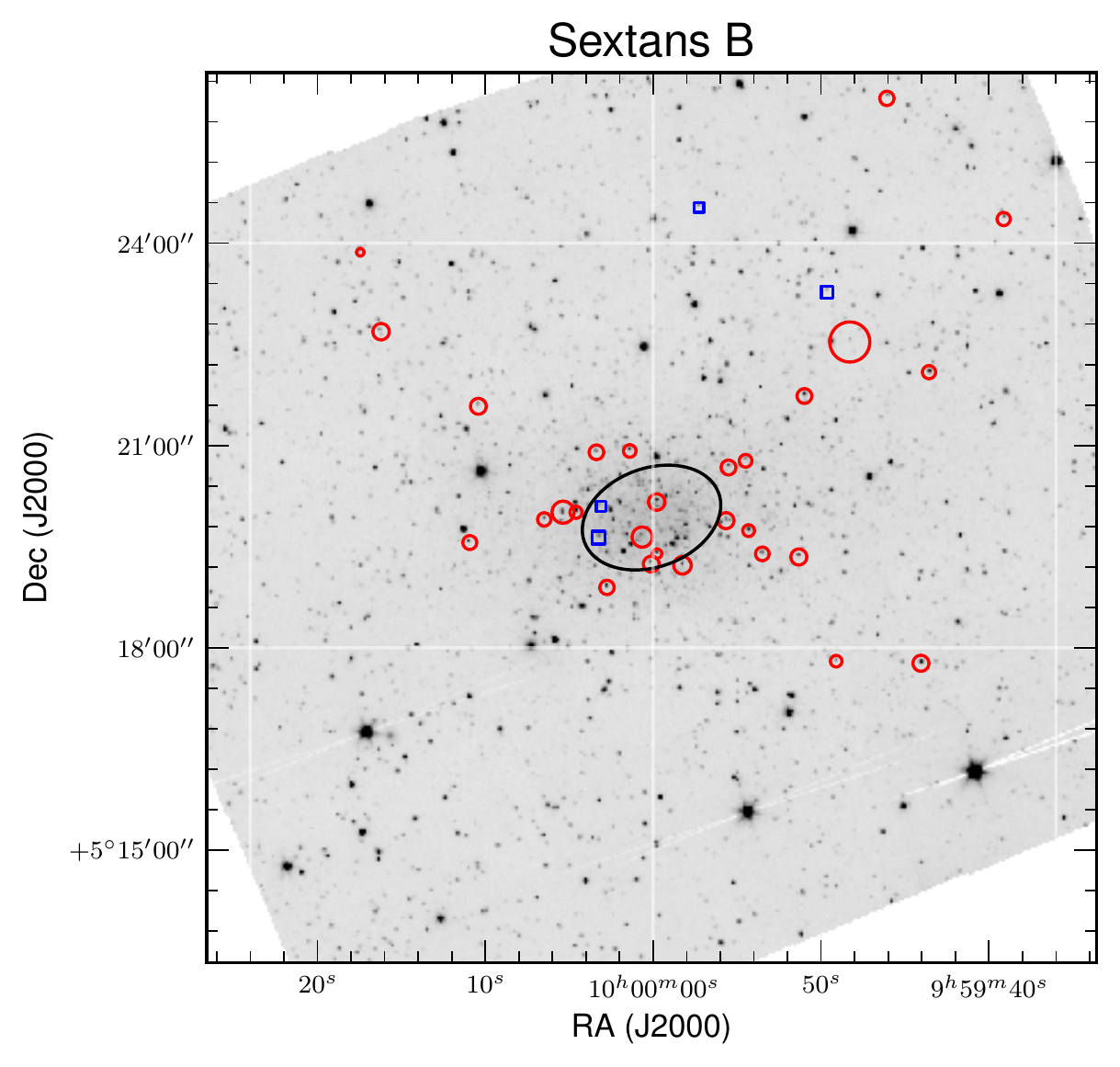}
}

\caption{Figure~\ref{fig:spat1} continued.}
\label{fig:spat2}
\end{figure}

\begin{figure}
\vbox{
\includegraphics[width=0.45\textwidth]{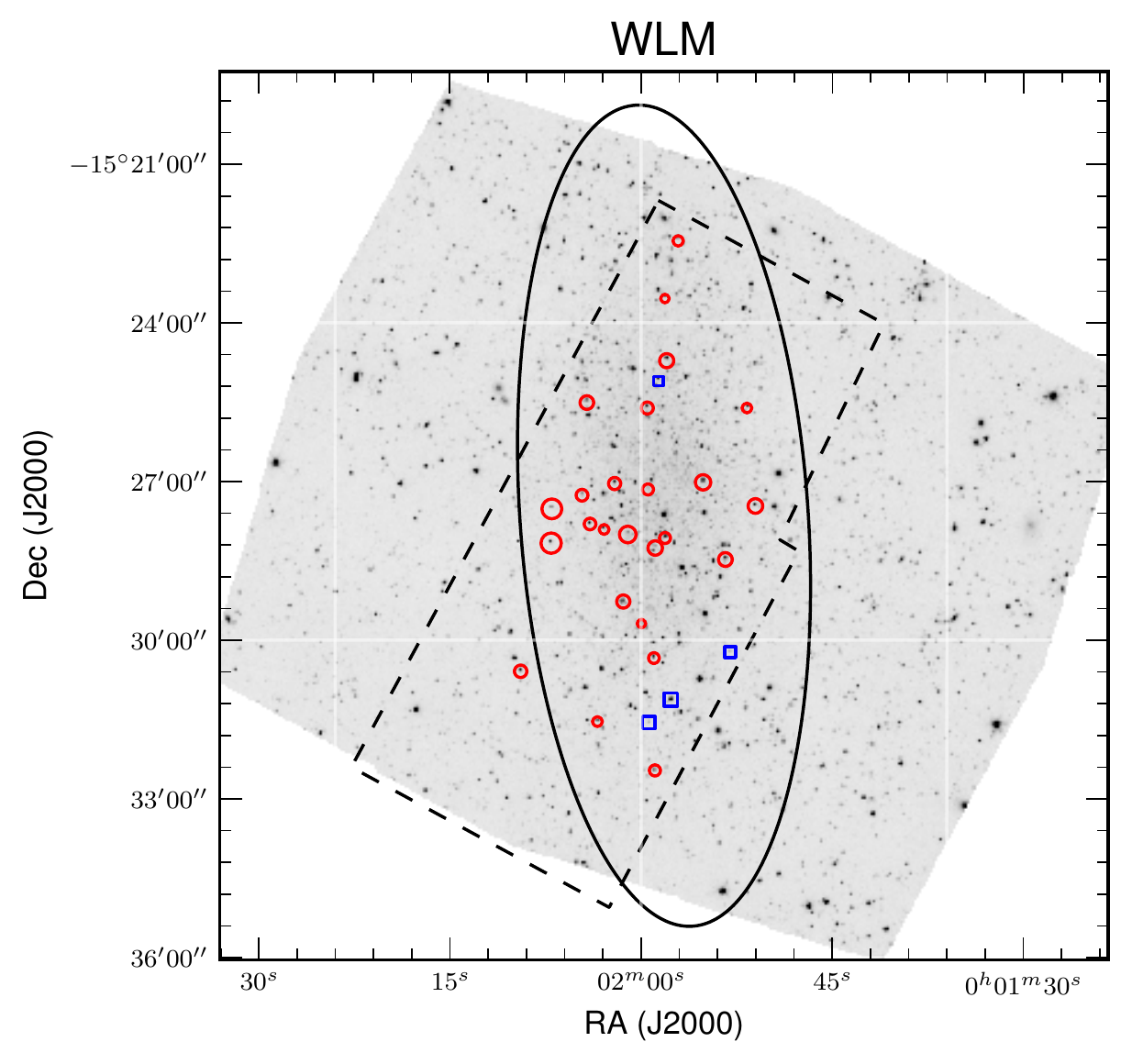}
}
\caption{Figure~\ref{fig:spat1} continued.}
\label{fig:spat3}
\end{figure}

Figures~\ref{fig:spat1}, \ref{fig:spat2} and \ref{fig:spat3} show the
{\it Spitzer} 3.6~\micron\ maps of galaxies with variable x-AGB
candidates.  

\end{document}